\documentclass[floats,floatfix,showpacs,amssymb,prd,twocolumn,superscriptaddress,nofootinbib,nolongbibliography,reprint]{revtex4-1}

\usepackage{amssymb,amsmath,verbatim,mathtools,needspace,enumitem,etoolbox,graphicx,physics,microtype,afterpage,xspace,tabularx,lmodern,multirow}
\usepackage{gensymb}
\usepackage{float}
\usepackage{breqn}
\usepackage[dvipsnames, usenames]{xcolor}
\definecolor{linkcolor}{rgb}{0.0,0.3,0.5}
\usepackage[unicode, colorlinks=true, linkcolor=linkcolor, citecolor=linkcolor, filecolor=linkcolor, urlcolor=linkcolor, linktocpage, breaklinks]{hyperref}
\usepackage[all]{hypcap}
\usepackage[T1]{fontenc}
\usepackage[utf8]{inputenc}
\usepackage[usenames,dvipsnames]{xcolor}
\hypersetup{colorlinks=true,citecolor=romared,linkcolor=romared,urlcolor=romared}

\definecolor{romared}{RGB}{142,0,28}

\newcommand{\be}{\begin{equation}}
\newcommand{\ee}{\end{equation}}

\def\be{\begin{equation}}
\def\ee{\end{equation}}
\newcommand{\beq}{\begin{eqnarray}}
\newcommand{\eeq}{\end{eqnarray}}

\usepackage{aas_macros}
\usepackage{makecell}
\usepackage{soul}
\usepackage{bm}
\usepackage{cleveref}

\usepackage{booktabs}
\usepackage{multirow}
\usepackage{lipsum}

\newcolumntype{Y}{>{\centering\arraybackslash}X}

\begin{document}
\title{Systematic bias from waveform modeling for binary black hole populations\\in next-generation gravitational wave detectors}

\begin{abstract}
Next-generation gravitational wave detectors such as the Einstein Telescope and Cosmic Explorer will have increased sensitivity and observing volumes, enabling unprecedented precision in parameter estimation. However, this enhanced precision could also reveal systematic biases arising from waveform modeling, which may impact astrophysical inference. We investigate the extent of these biases over a year-long observing run with $10^5$ simulated binary black hole sources using the linear signal approximation. To establish a conservative estimate, we sample binaries from a smoothed truncated power-law population model and compute systematic parameter biases between the \texttt{IMRPhenomXAS} and \texttt{IMRPhenomD} waveform models. For sources with signal-to-noise ratios above 100, we estimate statistically significant parameter biases in $\sim 3\%-20\%$ of the events, depending on the parameter. We find that the average mismatch between waveform models required to achieve a bias of $\leq 1\sigma$ for $99\%$ of detections with signal-to-noise ratios $\geq 100$ should be $\mathcal{O}(10^{-5})$, or at least one order of magnitude better than current levels of waveform accuracy.
\end{abstract}

\author{Veome Kapil}
\email{vkapil1@jhu.edu}
\affiliation{William H. Miller III Department of Physics and Astronomy\char`,{} Johns Hopkins University\char`,{} 3400 N. Charles Street\char`,{} Baltimore\char`,{} Maryland\char`,{} 21218\char`,{} USA}

\author{Luca Reali}
\email{lreali1@jhu.edu}
\affiliation{William H. Miller III Department of Physics and Astronomy\char`,{} Johns Hopkins University\char`,{} 3400 N. Charles Street\char`,{} Baltimore\char`,{} Maryland\char`,{} 21218\char`,{} USA}

\author{Roberto Cotesta}
\affiliation{William H. Miller III Department of Physics and Astronomy\char`,{} Johns Hopkins University\char`,{} 3400 N. Charles Street\char`,{} Baltimore\char`,{} Maryland\char`,{} 21218\char`,{} USA}

\author{Emanuele Berti}
\email{berti@jhu.edu}
\affiliation{William H. Miller III Department of Physics and Astronomy\char`,{} Johns Hopkins University\char`,{} 3400 N. Charles Street\char`,{} Baltimore\char`,{} Maryland\char`,{} 21218\char`,{} USA}

\date{\today}
\maketitle

\section{Introduction}

Gravitational wave (GW) astronomy is revolutionizing our understanding of relativistic astrophysics through the direct detection of double compact object (DCO) mergers. The LIGO-Virgo-KAGRA GW detector network has already observed $\mathcal{O}$(100) binary black hole (BBH) mergers~\cite{LIGOScientific:2021djp}, allowing us to chart the population of black holes in the local universe for the first time. The number of such observations is expected to increase dramatically with the introduction of next-generation (XG) GW detectors such as the Einstein Telescope (ET)~\cite{Punturo:2010zz} and Cosmic Explorer (CE)~\cite{Reitze:2019iox}.
The increased sensitivities and observing volumes of these detectors will open the door to unprecedented science, ranging from studies of DCO formation channels and multimessenger astronomy to probes of physics beyond general relativity, nuclear physics and dark matter~\citep{Maggiore:2019uih,Evans:2021gyd,Branchesi:2023mws,Evans:2023euw,Gupta:2023lga,Corsi:2024vvr}.

These applications will rely upon accurate measurements of the astrophysical parameters of DCOs, which in turn depend on the waveform models used to analyze the GW signals. Due to computational constraints, current waveform models are semianalytic approximations to two-body solutions in general relativity, which are then calibrated to numerical relativity (NR) simulations of DCO mergers. This gap between NR waveforms and semianalytic models can lead to systematic biases in the inferred binary parameters~\citep{Cutler:2007mi}. The problem of biased parameter inference is likely to be worse in regions of the parameter space where NR calibration points are sparse. 

The limited signal-to-noise ratio (SNR) achievable in current detectors leads to large enough statistical uncertainties on binary parameters that systematic biases are generally not significant.
At the sensitivity level of the third observing run of the LIGO-Virgo-KAGRA detectors (O3), the systematic biases from using inadequately calibrated waveform models are not particularly concerning (but see e.g.~\citep{Estelles:2021jnz,Hu:2022rjq} for cases in the O3 catalog where different waveform models produce inconsistent parameter posteriors).
Future ground-based GW detectors, however, are expected to observe DCO coalescences with SNRs in excess of $\mathcal{O}(100)$ or even $\mathcal{O}(1000)$~\citep{LIGOScientific:2016wof,Borhanian:2022czq}. The statistical uncertainties associated with such detections will be small enough that any systematic biases due to waveform errors could become significant. 

Some of the existing literature estimates the impact of waveform errors on a handful of loud binaries. Studies by \citet{Owen:2023mid} and \citet{Read:2023hkv}, among others, demonstrate the possibility of significant systematic biases due to waveform errors, even as early as O4. \citet{Purrer:2019jcp} compute the expected parameter biases for a few high-SNR binaries in CE/ET to be in excess of $100\sigma$. They show, albeit briefly, that waveform errors could also bias population model parameters. They further estimate that the mismatch between two given waveform models must be improved by $\sim 3$ orders of magnitude to ensure accurate parameter estimation for the loudest binaries in XG detectors. \citet{Hu:2022rjq} also conclude that a $3-4$ order of magnitude improvement in waveform accuracy would be required at the SNRs expected in future ground-based detectors.
Previous work has also shown that systematic bias will play a crucial role for the high-SNR massive black hole mergers observed by the Laser Interferometer Space Antenna (LISA)~\cite{Berti:2006ew,Cutler:2007mi,Pitte:2023ltw}, and that
parametrized tests of general relativity using high-SNR detections can produce biased results due to waveform modeling errors~\citep{Moore:2021eok, Saini:2022igm, Saini:2023rto, Hu:2022bji, DuttaRoy:2024aew}.

In this work, we expand and complement these studies by establishing a lower bound on the required waveform accuracy improvement for XG detectors. To this end, we consider the distribution of parameter biases across an astrophysical population of stellar mass BBHs. We first limit our focus to a pair of current waveform models from the same waveform family (\texttt{IMRPhenomD}~\cite{Husa:2015iqa, Khan:2015jqa} and \texttt{IMRPhenomXAS}~\cite{Pratten:2020fqn}), conservatively quantifying the impact that current waveform errors will have on a realistic population of XG GW observations. As a second step, we synthetically tune the accuracy of our chosen waveforms to model future improvements in waveform calibration. Using waveforms interpolated between \texttt{IMRPhenomD} and \texttt{IMRPhenomXAS}, we estimate a minimum accuracy requirement for future waveform models to produce consistent parameter estimation across an astrophysically motivated BBH population.

The plan of the paper is as follows.
In Sec.~\ref{sec:WFerrors} we discuss how the Fisher information matrix formalism can be used to estimate waveform errors and systematic biases through the formalism developed by Cutler and Vallisneri~\cite{Cutler:2007mi}, and we introduce our waveform interpolation scheme.
In Sec.~\ref{sec:populations} we present our model of the astrophysical BBH population.
In Sec.~\ref{section:wf_calibration_errors} we compute waveform errors within the Cutler-Vallisneri formalism, and estimate the waveform accuracy requirement for these systematic errors to be subdominant with respect to statistical errors in XG detectors.
In Sec.~\ref{sec:fullPE} we compare these results against full Bayesian parameter estimation calculations for selected binaries.
In Sec.~\ref{sec:discussion} we discuss the implications of our work and possible directions for future research.
In Appendix~\ref{app:network_cv_bias} we generalize the Cutler-Vallisneri calculation to a detector network,
in Appendix~\ref{app:alignment} we illustrate the importance of waveform alignment in the calculation of the biases on luminosity distance,
in Appendix~\ref{app:fref} we show how the choice of a reference frequency parameter affects the behavior of frequency-domain waveforms and of the bias, and in Appendix~\ref{app:interpolation} we show how the biases and statistical errors change as a function of the interpolation parameter between different waveforms.

\section{Statistical and Systematic Errors}
\label{sec:WFerrors}

In this section we discuss how to compute parameter errors and the Cutler-Vallisneri parameter bias within the Fisher information matrix formalism. We introduce the mismatch between waveforms, and we present a waveform interpolant that will be used in the rest of the paper to quantify how close two waveforms should be in order to achieve some desired maximum bias in the estimated parameters.

\subsection{Fisher information matrix}

The Fisher information matrix for a detected waveform $h(\vec \theta)$ which depends on a set of parameters $\vec \theta=\{\theta_i\}$ is defined as
\begin{equation}
    \Gamma_{ij} \equiv \left( \frac{\partial h}{\partial \theta^i} \middle \vert \frac{\partial h}{\partial \theta^j} \right),
\end{equation}
where the inner product between two waveforms in the frequency domain is defined as
\begin{equation}
    ({a}|{b}) \equiv 4 \Re \int_{f_{\rm min}}^{f_{\rm max}} \frac{\widetilde{a}(f) \cdot \widetilde{b}(f)^*}{S_n(f)} \text{d}f,
\end{equation}
and $S_n(f)$ is the noise power spectral density for the given detector. For a network of detectors, the network Fisher matrix is given by
\begin{equation}
    \Gamma_{\rm Net} = \Sigma_{{\rm D}=1}^{\rm N_{\rm D}} \Gamma_{\rm D},
    \label{eq:fisher_network}
\end{equation}
where the subscript $\rm D$ labels each detector in a network made up of $\rm N_D$ total detectors. In the linear signal approximation and under the assumption of Gaussian posteriors~\citep{Vallisneri:2007ev}, the network's covariance matrix is
\begin{equation}
    \Sigma_{\rm Net} = (\Gamma_{\rm Net})^{-1}.
\end{equation}
The variances and covariances for each parameter are given by the diagonal and off-diagonal elements of $\Sigma_{\rm Net}$, respectively, such that
\begin{equation}
    \sigma_{\theta_i} = \sqrt{\Sigma_{ii, \rm{Net}}}.
\end{equation}

We use the \texttt{GWBENCH}~\citep{Borhanian:2020ypi} code to compute Fisher information matrices and waveform derivatives for our simulated BBH population. The detector network consists of one 40 km Cosmic Explorer detector in Idaho, USA and one 20 km Cosmic Explorer detector in New South Wales, Australia, plus one Einstein Telescope detector at design sensitivity in Cascina, Italy (from now on, for brevity, a ``2CE+ET'' network). The specific locations and orientations of these detectors can be found in Table~III of Ref.~\cite{Borhanian:2020ypi} under the labels \texttt{C}, \texttt{S}, and \texttt{E}, respectively. The power spectral densities (PSDs) adopted here correspond to the 4020ET network of Ref.~\citep{Gupta:2023lga}.

We set the minimum frequency for the detector response to be 5 Hz, which is also the reference frequency used to generate the waveforms.

We compute the waveform derivatives and Fisher matrices for the following parameters: chirp mass $\mathcal{M}_{\rm c}$, symmetric mass ratio $\eta$, components of the primary and secondary dimensionless spins along the orbital angular momentum $\chi_{1, \rm{z}}$ and $\chi_{2, \rm{z}}$, luminosity distance $D_{\rm L}$, coalescence time $t_{\rm c}$, coalescence phase $\phi_{\rm c}$, binary inclination $\iota$, right ascension RA, declination DEC, and polarization angle $\psi$. These parameters encapsulate all the degrees of freedom in the waveform models considered for this study.

Although we compute derivatives with respect to all of these parameters, in this study we only consider biases on the intrinsic binary parameters and on the luminosity distance $D_{\rm L}$. Our tests show that the posteriors for the intrinsic parameters are generally well approximated by Gaussian distributions, and so the Fisher formalism is better suited to these parameters.
To incorporate physical bounds on the statistical errors (particularly for the spins) within the Fisher matrix formalism, we crudely impose prior bounds following Refs.~\cite{Cutler:1994ys,Poisson:1995ef,Berti:2004bd}, i.e., we rewrite the Fisher matrix for each binary as
\begin{equation}
  \label{eq:prior}
    \Gamma_{\rm bounded} = \Gamma+\Gamma^{(0)}.
\end{equation}
Here, $\Gamma$ is the na\"ive output of the Fisher matrix computation, and $\Gamma^{(0)}$ is the Fisher matrix corresponding to the multivariate Gaussian prior imposed on the system. The covariance matrix for the system can then be written as 
\begin{equation}
    \Sigma_{\rm bounded} = (\Gamma+\Gamma^{(0)})^{-1}.
\end{equation}
For simplicity, we only impose priors on $\chi_{1, \rm z}$ and $\chi_{2, \rm z}$ such that their statistical uncertainties are bounded in the range $[-1,\,1]$. We find that imposing priors on other parameters (such as $\mathcal{M}_{\rm c}$ and $\eta$) does not significantly impact the covariance, since the na\"ive error estimates for those parameters are generally small compared to the prior range.

\subsection{Cutler-Vallisneri bias}
\label{subsection:cv_bias}

Consider the scenario where a DCO merger produces some true gravitational waveform $h_{\rm TR}$, which we might detect with our network of ground-based GW detectors. Let us denote by $h_{\rm AP}$ some hypothetical state-of-the-art waveform model approximant used to infer the binary parameters. Due to imperfect calibration of the approximate waveform model, there will be regions of the parameter space $\vec{\theta}$ such that $h_{\rm TR}(\vec{\theta}) \neq h_{\rm AP}(\vec{\theta})$. The resulting systematic bias on the parameter $\theta^i$ is given by
\begin{equation}
  \Delta \theta^i =  (\Gamma_{\rm AP, Net}^{-1})^{ij} \, \Sigma_{{\rm D}=1}^{\rm N_{\rm D}} \, (\partial_j (h_{\rm AP})_{\rm D} | (h_{\rm TR})_{\rm D} - (h_{\rm AP})_{\rm D}),
  \label{eq:cv_bias}
\end{equation}
where $\Gamma_{\rm AP, Net}$ is the network Fisher matrix defined in Eq.~\eqref{eq:fisher_network} evaluated using the $h_{\rm AP}$ waveform, and $(h_{\rm AP})_{\rm D}$ and $(h_{\rm TR})_{\rm D}$ represent the responses of the approximate and true waveforms in detector $\rm D$, respectively.

To our knowledge, this bias was first computed in Refs.~\cite{Flanagan:1997kp,Cutler:2007mi}. The generalization to a detector network given in Eq.~\eqref{eq:cv_bias} above can be found in Appendix~\ref{app:network_cv_bias}.
From now on, we will refer to $\Delta \theta^i$ as the Cutler-Vallisneri bias.

In practice, we do not have access to a perfect model for the ``true'' waveform $h_{\rm TR}$. To emulate the effect of waveform modeling errors, we thus use two waveform models from the \texttt{IMRPhenom} family of phenomenological GW approximants. Specifically, we use \texttt{IMRPhenomXAS}~\cite{Pratten:2020fqn} to represent the ``true'' waveform $h_{\rm TR}$, and \texttt{IMRPhenomD}~\cite{Husa:2015iqa, Khan:2015jqa} as the imperfectly calibrated waveform model $h_{\rm AP}$.

These are both frequency-domain models of the gravitational radiation from nonprecessing BBHs which include only the dominant $(2,\,2)$ spherical harmonic mode.
\texttt{IMRPhenomXAS} is an update to \texttt{IMRPhenomD} with better calibration to reference waveforms with large mass ratios and unequal spins~\citep{Pratten:2020fqn}.  As a result, any differences between \texttt{IMRPhenomXAS} and \texttt{IMRPhenomD} should be largest for volumes in the parameter space where the fewest calibration waveforms exist. We use semianalytic waveform models, rather than NR waveforms, to represent the true as well as the approximate waveforms so that we can study waveform errors for a population of BBHs spanning a large volume of the parameter space. Since our goal is to estimate waveform accuracy at the order-of-magnitude level, we deem these waveform models sufficient for this study. We expect waveform errors to be larger between models from different families and with different physics, so our study represents a conservative estimate of the resulting biases. We leave more detailed comparisons with other waveform models to future work.

\subsection{Waveform mismatch}
The differences in waveform models can be quantified by the relative disagreement, or mismatch, between the waveforms at a given point in parameter space. The mismatch $\mathcal{M}$ between an approximate waveform $h_{\rm AP}$ and the true waveform $h_{\rm TR}$ can be computed as
\begin{equation}
\label{eq:mismatch}
\mathcal{M} = 1 - \max_{\phi_{\rm c},t_{\rm c},\psi} \frac{({h}_{\rm AP}(\bm{\vec{\theta}}, \phi_{\rm c}, t_{\rm c}, \psi) \, | \, {h}_{\rm TR})}{\sqrt{({h}_{\rm AP} | {h}_{\rm AP})} \sqrt{({h}_{\rm TR} | {h}_{\rm TR})}},
\end{equation}
where we maximize over the parameters that capture differences in waveform convention. A mismatch of zero corresponds to two exactly equivalent waveforms, while a mismatch of unity indicates maximally distinct waveforms in the frequency domain.

\subsection{Interpolating waveform models}
\label{subsection:hybrid_wf}
We are interested in not only quantifying systematic biases between current waveform models, but also estimating the requirements on future waveform models to mitigate these biases. To find these accuracy requirements in practice, we would like to generate waveforms with arbitrary levels of calibration with respect to a reference waveform model. To accomplish this, we first decompose our true and approximate waveforms into an amplitude and a phase:
\begin{align}
    h_{\rm TR}(f) &= A_{\rm TR}(f) \exp[{i \, \phi_{\rm TR}(f)}], \\
    h_{\rm AP}(f) &= A_{\rm AP}(f) \exp[{i \, \phi_{\rm AP}(f)}].
\end{align} 
We can now construct a new waveform
\begin{equation}
    h_{\rm AP}'(f) = A_{\rm AP}'(f) \exp[i \, \phi_{\rm AP}'(f)],
\end{equation}
where the amplitude and phase are obtained by linearly interpolating between the two original waveforms as follows:
\begin{align}
    \label{eq:hybrid_wf_amp}
    A_{\rm AP}' &= \lambda\, A_{\rm TR}+ (1-\lambda)\, A_{\rm AP} , \\
    \label{eq:hybrid_wf_phase}
    \phi_{\rm AP}' &= \lambda\, \phi_{\rm TR} + (1-\lambda)\, \phi_{\rm AP} .
\end{align}
This approach allows us to place waveforms anywhere between two given waveform models, and thus to emulate future levels of waveform mismatch. By construction, the $h_{\rm AP}'$ waveform is expected to be in better agreement with $h_{\rm TR}$ than the $h_{\rm AP}$ waveform (see Appendix~\ref{app:interpolation}). The parameter $\lambda$ is best interpreted as a perturbation around $h_{\rm TR}(f)$, in the direction of $h_{\rm AP}(f)$, such that $|1 - \lambda| \ll 0$. We avoid larger values of $\lambda$, which may lead to unphysical parameter biasing artifacts.

\section{Astrophysical Population of Binary Black Holes}
\label{sec:populations}

The accuracy of available waveform models can vary drastically as a function of the intrinsic parameters of the binary (such as masses and spins), depending on the availability and accuracy of NR calibration waveforms. One of our goals is to understand what fraction of the BBH systems observable with XG GW detectors would be strongly impacted by these biases. To answer this question, we will quantify the aggregate severity of waveform calibration biases over simulated binary populations compatible with current LIGO-Virgo-KAGRA observations.
Our results are naturally sensitive to the chosen population model, so it will be worth expanding the present analysis using various parametric and nonparametric populations in the future. Following Ref.~\cite{Evans:2021gyd}, we assume that our detector network will detect $10^5$ BBH sources over a one year-long observing run. We sample the masses, spins, orientations, and sky locations of these BBHs from probability density functions consistent with our astrophysical assumptions. While the specific population model chosen for our study may not precisely represent the true distribution of BBHs in the Universe, it should provide a reasonable first estimate of the distribution of systematic biases for some of the most interesting binary parameters.

\noindent
{\bf \em Masses.} The mass distribution of our BBH population is chosen to follow the \textsc{Truncated} model of Ref.~\citep{LIGOScientific:2021psn}, with some modifications at the low-mass end.
For our population of $10^5$ BBH systems, the primary mass $m_1$ is drawn from a modified truncated power-law distribution
\begin{multline}
    p(m_1| \alpha, \eta, m_0, m_{\rm min}, m_{\rm max}) \propto \\
    m_1^\alpha \times h(m_1|m_0, \eta) \times \mathcal{B}(m_1 | m_{\rm min}, m_{\rm max}). \label{eq:primary_mass_dist}
\end{multline}
The proportionality sign indicates that the final probability density function should be normalized, but we omit the full expression here for clarity. The first term of Eq.~\eqref{eq:primary_mass_dist} represents the usual power law with exponent $\alpha$. The next term is a high-pass Butterworth filter with roll-off mass $m_0 = 5 \rm{M}_\odot$ and sharpness parameter $\eta = 50$, given by
\begin{equation}
    h(m_1|m_0,\eta) = \left[1+ \left(\frac{m_0}{m_1}\right)^\eta \right]^{-1}.
\end{equation}
This term is responsible for smoothing the lower end of the primary mass distribution so that it remains continuous. The last term is the boxcar function, used to enforce that $m_{1}$ is between the limits $m_{\rm min}$ and $m_{\rm max}$, and given by
\begin{equation}
    \mathcal{B}(m_1 | m_{\rm min}, m_{\rm max}) = 
    \begin{cases}
        1 & \text{if } m_1 \in [m_{\rm min}, m_{\rm max}],\\
        0 & \text{otherwise}.
    \end{cases}
\end{equation}
In our simulations, we fix $m_{\rm min} = 3 \rm{M}_\odot$ and $m_{\rm max}=60 \rm{M}_\odot$. The Butterworth filter and power law ensure that the peak of the primary mass distribution is at $5 \rm{M}_\odot$. The secondary mass $m_2$ is then drawn uniformly between $m_{\rm min}$ and $m_1$.

When computing the waveform responses and derivatives, these masses are converted to chirp mass
$\mathcal{M}_{\rm c} = (m_1 m_2)^{3/5}(m_1 + m_2)^{-1/5}$
and symmetric mass ratio
$\eta = (m_1 m_2)(m_1 + m_2)^{-2}$.

Due to the gravitational redshift, only the redshifted masses of a binary are observable via GWs. To properly emulate GW detections, the binary component masses sampled above must be converted into detector-frame redshifted masses such that
$m_{\rm det} = m (1+z)$,
where the redshift for each binary is sampled from a suitable probability distribution. Similarly, we have
$\mathcal{M}_{\rm{c, det}} = \mathcal{M}_{\rm c} (1+z)$.

\noindent
{\bf \em Spins.} The spin magnitude for each binary component $\chi_i$ is independently drawn from a beta distribution
\begin{equation}
    p(\chi_i| a,b) \propto \chi_i ^{a-1} (1-\chi_i)^{b-1}.
\end{equation}
We choose $a=2$ and $b=7$, which generally reproduces the \textsc{Default} spin model from Ref.~\cite{LIGOScientific:2021psn}.
The spin orientations are isotropically distributed on the sphere, such that the polar angle $\theta$ obeys $p(\cos \theta) =   U [-1, 1]$ and the azimuthal angle $\phi$ is distributed as $p(\phi) =  U [0, 2 \pi]$. We limit our study to waveform models with aligned spins, so we consider only the $\chi_{i,z} = \chi_i \cos \theta$ component.

\noindent
{\bf \em Sky position and orientation.}
The right ascension and declination of the binaries are sampled uniformly on the sphere: 
\begin{align}
    p(\rm RA) &\propto U[0, 2 \pi], \\
    p(\cos(\Theta)) &\propto U[-1, 1],
\end{align}
where $\Theta = \rm{DEC}$$+ \pi/2$. 
The inclination angle $\iota$ is sampled uniformly in $\cos \iota$, where an angle of $0$ ($\pi/2$) represents a face-on (edge-on) binary.
The polarization angle $\psi$ is sampled uniformly between $0$ and $2\pi$.

\noindent
{\bf \em Coalescence time and phase.}
The coalescence time, $t_c$, and phase, $\phi_c$, are arbitrarily set to zero for all binaries in the population when evaluating $h_{\rm TR}$. These parameters are extrinsic, and can be chosen freely for each waveform model such that the overlap with the GW signal is maximized~\citep{Dhani:2024jja, Allen:2005fk, Buonanno:2009zt}. We therefore evaluate $h_{\rm AP}$ at the values of $t_c$ and $\phi_c$ that minimize the mismatch defined in Eq.~\eqref{eq:mismatch}, to allow for differences in conventions between the two waveform models.
We became aware of the importance of aligning the waveforms when computing the bias formula from the authors of Ref.~\citep{Dhani:2024jja}. 
We therefore evaluate $h_{\rm AP}$ at the values of $t_c$ and $\phi_c$ that minimize the mismatch defined in Eq.~\eqref{eq:mismatch}, to allow for differences in conventions between the two waveform models. In practice, aligning \texttt{IMRPhenomD} and \texttt{IMRPhenomXAS} in coalescence phase and time can lower the computed biases by several orders of magnitude, as first derived by~\citet{Dhani:2024jja}.

\noindent
{\bf \em Redshift.}
The source redshifts are drawn from the \citet{Madau:2016jbv} probability distribution, given by 
\begin{equation}
    \Psi(z) = 0.01 \frac{(1+z)^{2.6}}{1 + [(1+z)/3.2]^{6.2}} \rm{M}_\odot \rm yr^{-1} \rm Mpc^{-3}.
\end{equation}
The range of possible redshifts is chosen to be $z \in [0.02, 50]$ to match the population in Ref.~\citep{Borhanian:2022czq}. Wherever necessary, the redshift is converted to $D_{\rm L}$ (and vice versa) using the Planck18~\citep{Planck:2018vyg} cosmology. 
For simplicity, we assume that the star formation redshift distribution is the same as the merger redshift distribution, i.e., we do not explicitly model the delay times. The effect of this assumption will likely be to underestimate the significance of waveform biases, since a more careful modeling of delay times will cause BBHs to merge at lower redshifts, and thus produce smaller statistical errors on average.

\begin{figure}[t]
\includegraphics[trim={0.5cm 1cm 0 0}, 
width=\linewidth]{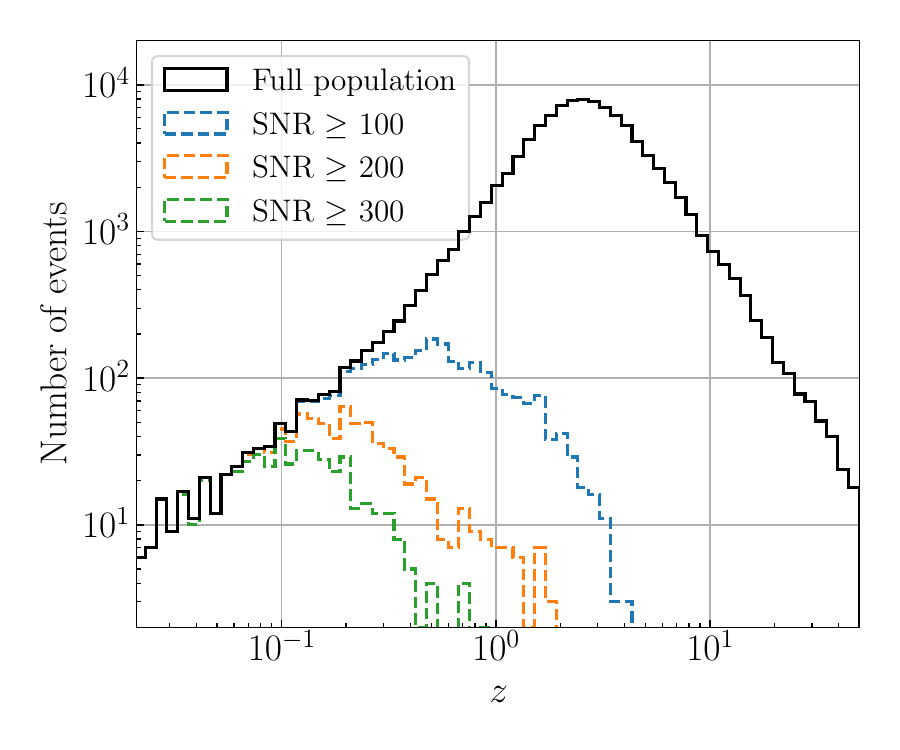}
\caption{Redshift distributions of the binaries observed at various SNR thresholds by a 2CE+ET network, as sampled from the \citet{Madau:2016jbv} distribution.}
\label{fig:snr_vs_z}
\end{figure}

In Fig.~\ref{fig:snr_vs_z} we show the redshift distribution of the simulated observable population at various SNR thresholds. The chosen detector network can detect simulated BBHs with SNR of $\mathcal{O}(100)$ out to $z \sim 3$. The median SNR across the entire population is about $22$, and the maximum SNR for a binary in our population is $\sim 4000$. This SNR distribution is in reasonable agreement with recent forecasts of the capabilities of XG detectors~\citep{Muttoni:2023prw,Pieroni:2022bbh,Iacovelli:2022bbs}.

\section{Results}
\label{section:wf_calibration_errors}

In our estimates of statistical and systematic parameter estimation errors, we will limit our attention to the loudest BBH sources, i.e., those with SNR $\geq 100$.
These systems (roughly 3000 binaries in our population model) are particularly valuable for astrophysical applications due to the small parameter uncertainties, and they are also most affected by any systematic biases between different waveform models.
The event posteriors of high-SNR BBHs are also generally well approximated by Gaussian distributions, and so the Fisher information formalism used in this work is expected to be a good approximation for these events. For the same reason, we also expect the systematic bias computed under the linear signal approximation to be reliable for these binaries. Wherever necessary, we impose a limit in postprocessing to ensure that the biases do not exceed physical bounds.

\begin{figure}[t]
\includegraphics[trim={1cm 0 0 0}, 
width=\linewidth]{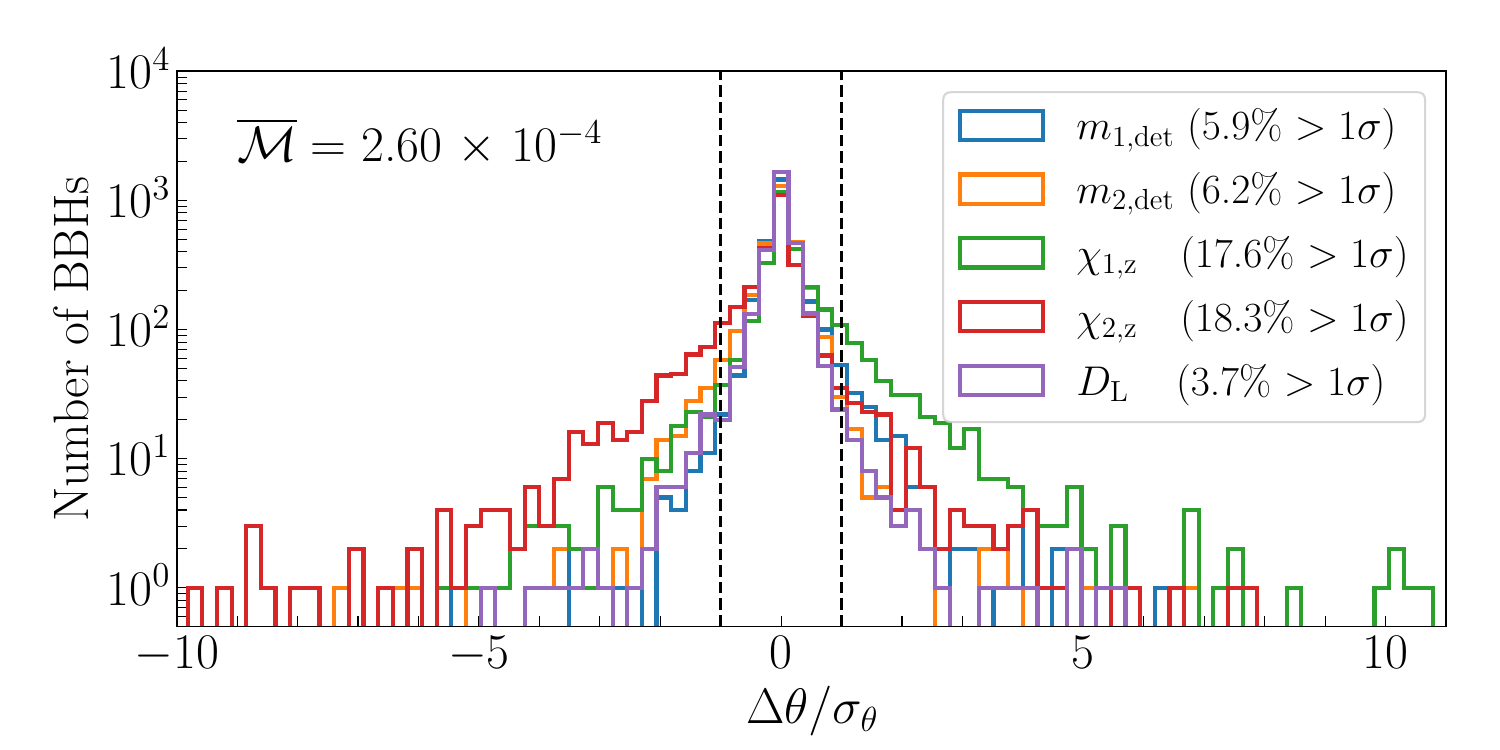}
\caption{Significance of waveform calibration biases between \texttt{IMRPhenomD} and \texttt{IMRPhenomXAS}, for various parameters, over the SNR $\geq$ 100 subpopulation. All the biases have been normalized by the statistical uncertainties, such that the dashed lines indicate the $1\sigma$ bias thresholds. The average mismatch between the true and approximate waveforms computed across this set of binaries is shown in the top left corner, and the percentages of events with $> 1\sigma$ bias are listed in the legend.}
\label{fig:phenomd_vs_xas_bias_hist}
\end{figure}

\begin{figure}[t]
  \centering
  \begin{tabular}{@{}c@{}}
    \includegraphics[trim={0.8cm 0 0 0},width=\linewidth]{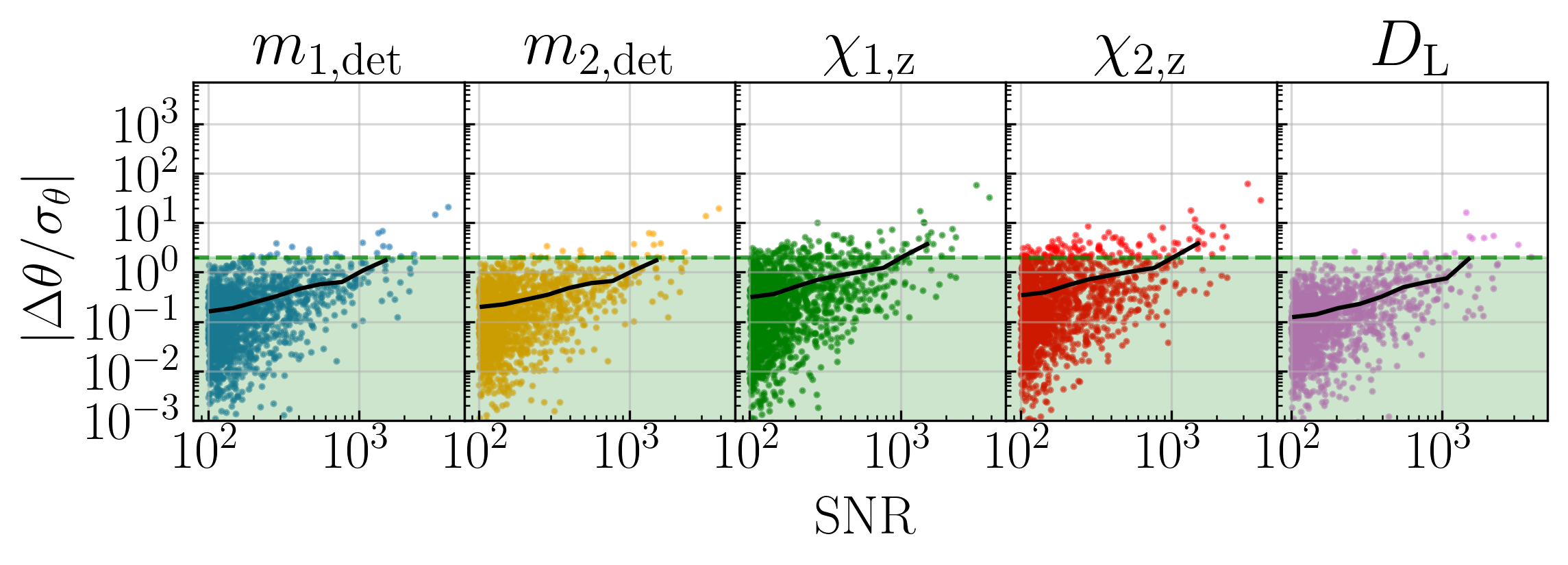} 
  \end{tabular}
  \begin{tabular}{@{}c@{}}
    \includegraphics[trim={0.8cm 0 0 0},width=\linewidth]{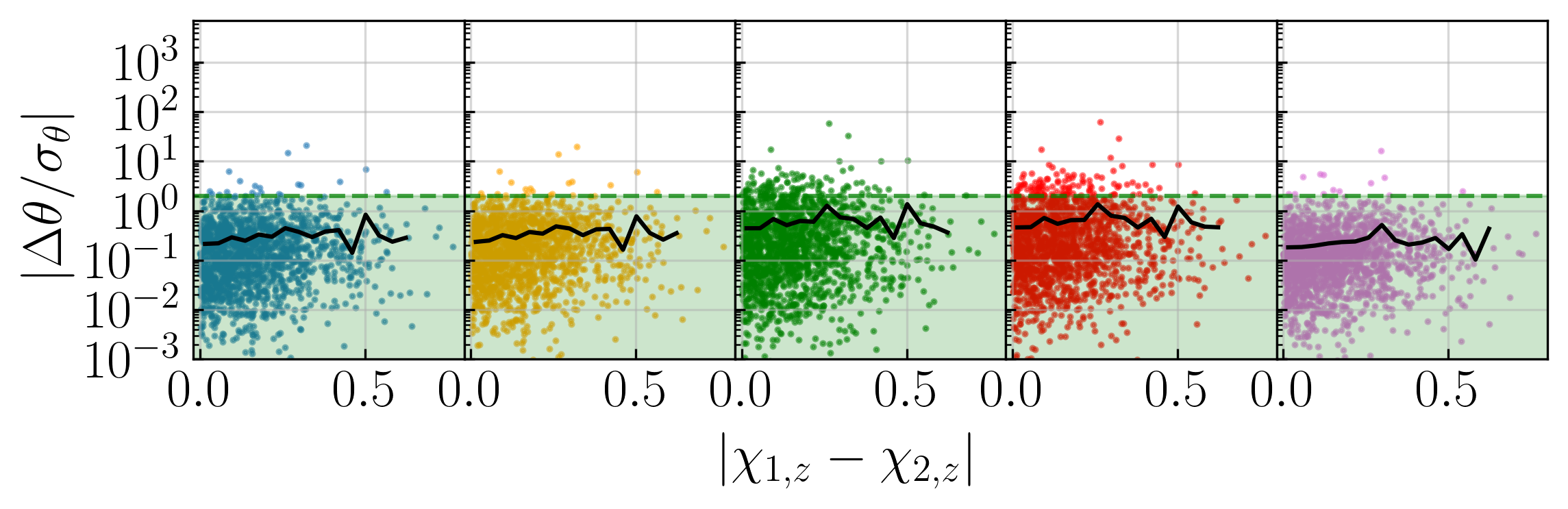} 
  \end{tabular}
  \begin{tabular}{@{}c@{}}
    \includegraphics[trim={0.8cm 0 0 0},width=\linewidth]{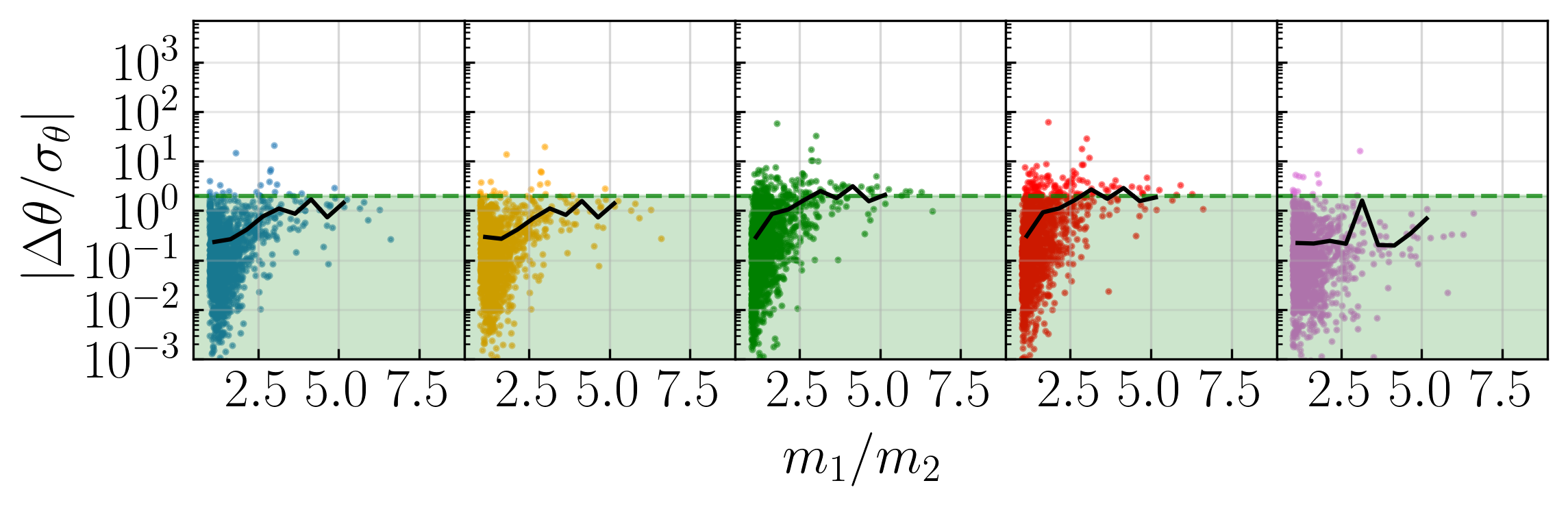} 
  \end{tabular}
  \begin{tabular}{@{}c@{}}
    \includegraphics[trim={0.8cm 0 0 0},width=\linewidth]{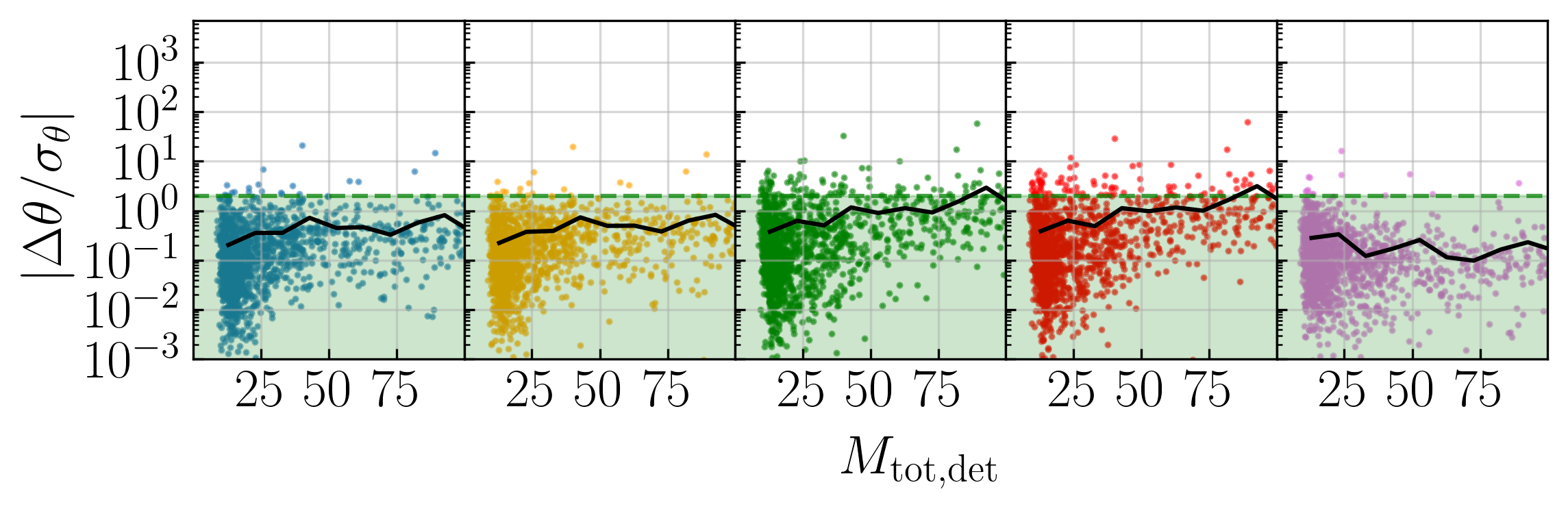} 
  \end{tabular}
  \caption{The absolute value of Cutler-Vallisneri biases for various parameters vs SNR (top), absolute difference between spins (2nd from top), mass ratio (3rd from top), and detector-frame total mass (bottom). Each dot is one binary from the SNR~$\geq$100 subpopulation, and the solid black line tracks the rolling average of the bias for each parameter. Biases are normalized to the statistical errors for each binary. The green shaded region represents biases below $1\sigma$.}\label{fig:bias_vs_params}
\end{figure}

\subsection{Current waveform model accuracy}

\begin{table*}
    \centering
   \renewcommand{\arraystretch}{1.5}
    \begin{tabular}{ccccc}
        \hline
        \multirow{2}{*}{Parameter} & \multicolumn{3}{c}{Median Absolute Bias $\overline{|\Delta \theta|}$ ($\overline{|{\Delta \theta/\sigma_{\theta}}|}$)} & \multirow{2}{*}{SNR$_{1\sigma}$}\\ \cline{2-4} &
        SNR $\geq$ 100 & SNR $\geq$ 200 & SNR $\geq$ 300 \\
        \hline 
        \hline
        $m_{1, \rm det}$ & 0.24 M$_{\odot}$ (0.1$\sigma$) & 0.25 M$_{\odot}$ (0.2$\sigma$) & 0.26 M$_{\odot}$ (0.4$\sigma$) & {$\sim$500*}\\
        $m_{2, \rm det}$ & 0.18 M$_{\odot}$ (0.1$\sigma$) & 0.18 M$_{\odot}$ (0.3$\sigma$) & 0.21 M$_{\odot}$ (0.4$\sigma$) & {$\sim$500*}\\
        $m_{1}$ & 0.18 M$_{\odot}$ (0.1$\sigma$) & 0.22 M$_{\odot}$ (0.3$\sigma$) & 0.25 M$_{\odot}$ (0.4$\sigma$) & {$\sim$500*}\\
        $m_{2}$ & 0.14 M$_{\odot}$ (0.1$\sigma$) & 0.16 M$_{\odot}$ (0.3$\sigma$) & 0.19 M$_{\odot}$ (0.4$\sigma$) & {$\sim$500*}\\
        $\chi_{1, \rm z}$  & 0.07 ($0.2\sigma$) & 0.12 ($0.4\sigma$) & 0.14 ($0.5\sigma$) & {$\sim$400}\\
        $\chi_{2, \rm z}$ & 0.10 ($0.2\sigma$) & 0.16 ($0.4\sigma$) & 0.19 ($0.6\sigma$) & {$\sim$400}\\
        $D_{\rm L}$ & 12.7 Mpc ($0.1\sigma$) & 3.4 Mpc ($0.2\sigma$) & 2.2 Mpc ($0.3\sigma$) & {$\sim$700*}\\
        $z$ & $1.9\times 10^{-3} (0.1\sigma)$ & $6.1\times 10^{-4} (0.2\sigma)$ & $4.4\times 10^{-3} (0.3\sigma)$ & {$\sim$700*}\\
        \hline
    \end{tabular}
\caption{Median magnitudes of the Cutler-Vallisneri biases, $\overline{|\Delta \theta|}$ (and their significance with respect to statistical errors, $\overline{|{\Delta \theta/\sigma_{\theta}}|}$), for subpopulations above various SNR thresholds in the 2CE+ET detector configuration. SNR$_{1\sigma}$ denotes the approximate SNR threshold above which the median significance of the bias in each parameter is expected to exceed unity. For parameters that never exceed a median bias of $1\sigma$ in our population due to limited statistics, the SNR$_{1\sigma}$ thresholds are calculated by extrapolation, and are marked with an asterisk (*). \label{tab:med_bias_vs_snr}}
\end{table*}

In Fig.~\ref{fig:phenomd_vs_xas_bias_hist} we show the distribution of Cutler-Vallisneri biases between \texttt{IMRPhenomXAS} and \texttt{IMRPhenomD}, $\Delta \theta$, normalized to the statistical uncertainties obtained from the Fisher analysis, $\sigma_\theta$, for the following binary parameters: $m_{\rm 1, det}$, $m_{\rm 2, det}$, $\chi_{\rm 1, z}$, $\chi_{\rm 2, z}$, and $D_{\rm L}$. Note that while both $h_{\rm TR}$ and $h_{\rm AP}$ enter the calculation of $\Delta\theta$ through the inner product of Eq.~\eqref{eq:cv_bias}, the Fisher matrix is always computed only in terms of the recovered signal $h_{\rm AP}$.

As a rule of thumb, we will say that a parameter is unbiased (biased) if $|\Delta \theta/\sigma_\theta|\leq 1$ ($|\Delta \theta/\sigma_\theta|>1$, respectively).
The differences between \texttt{IMRPhenomXAS} and \texttt{IMRPhenomD} can bias parameter values by at least $1\sigma$ for a few percent of this high-SNR population, with spin being the most affected parameter.

Although not shown in this figure (but see Fig.~\ref{fig:bias_vs_params}), the largest biases can reach $\geq 20 \sigma$ for the masses, and $\geq 50 \sigma$ for the spins. Regardless, there are at least 2500 binaries in our population with SNR~$\geq~100$ and unbiased intrinsic parameters.
The average mismatch between the two waveform models across the SNR~$\geq~100$ binaries is $\overline{\mathcal{M}} = 2.6 \times 10^{-4}$. Note that the parameters shown can be biased in either direction, depending on the specific configuration of the binary system. As such, the average values of the biases are consistent with zero for most parameters when we consider the entire population.

In Fig.~\ref{fig:bias_vs_params} we show a few alternate visualizations of the magnitudes of the normalized biases $|\Delta \theta/\sigma_\theta|$ for various binary parameters.

The top panel of this plot displays a strong positive correlation between the SNR of a binary and the significance of the waveform biases. This is expected, as the statistical uncertainties scale as $\rm{SNR}^{-1}$, while it can be shown that the systematic uncertainties from Eq.~\eqref{eq:cv_bias} do not depend on SNR.

The spin of a binary also seems to have some impact on the magnitude of the biases: the second panel from the top in Fig.~\ref{fig:bias_vs_params} shows that large differences in spin magnitudes are weakly correlated with larger biases. Although this is not a particularly strong effect in the data, it aligns with the differences in calibration between \texttt{IMRPhenomD} and \texttt{IMRPhenomXAS}, with the former waveform model being calibrated primarily to equal-spin binary simulations~\cite{Khan:2015jqa}, and the latter being calibrated to more unequal-spin simulations~\cite{Pratten:2020fqn}. As the spin parameter space gets explored better in simulations, differences in calibration between different waveform models will likely become more apparent.

The significance of Cutler-Vallisneri biases for intrinsic parameters is also somewhat positively correlated to the binary mass ratio $m_1/m_2$ (third panel) and to the total detector-frame mass (bottom panel). Compared to \texttt{IMRPhenomD}, the \texttt{IMRPhenomXAS} waveform model is calibrated to additional waveforms with large mass ratios. Additionally, \texttt{IMRPhenomD} exhibits worse calibration to high-mass binaries than to low-mass binaries (see Fig.~15 of~\cite{Pratten:2020fqn}), which may explain the trends in Fig.~\ref{fig:bias_vs_params}. Although not explored further in this study, other population models allowed by the Gravitational Wave Transient Catalog 3 (GWTC-3)~\citep{KAGRA:2021duu} exhibit overdensities in the primary mass spectrum around $m_{1}\approx 10 M_{\odot}$ or higher. Our assumed population is dominated by lower mass black holes, and hence, our study is conservative in terms of the assumed mass distribution. The median biases for the true observed BBH population will likely be larger than our estimates. 

In Table~\ref{tab:med_bias_vs_snr}, we show the behavior of the median magnitude of Cutler-Vallisneri biases for our population of BBHs for subpopulations with various SNR thresholds, with predictions for the threshold above which the median biases would overcome statistical uncertainties. For each parameter in the table, we present the median absolute values of Cutler-Vallisneri biases, $\overline{|\Delta \theta|}$, for three sets of SNR cutoffs. The SNR~$\geq 100$ threshold is the fiducial value adopted for Fig.~\ref{fig:phenomd_vs_xas_bias_hist}. 

The absolute parameter biases computed via the Cutler-Vallisneri formalism are independent of SNR, and so these values are largely constant across SNR thresholds. In parentheses, we indicate how significant the biases are, normalized by the statistical uncertainty in each of the parameters ($\overline{|{\Delta \theta/\sigma_{\theta}}|}$). Since $\sigma_{\theta}$ scales inversely with SNR, we find that the significance of the bias increases almost linearly with the SNR cutoff. 
The final column of Table~\ref{tab:med_bias_vs_snr} lists the SNR threshold, SNR$_{1 \sigma}$, above which the median bias is expected to exceed $1\sigma$. Due to lack of sufficient statistics at higher SNR thresholds, we do not always find subpopulations with $\overline{|{\Delta \theta/\sigma_{\theta}}|}\geq 1$ for the listed parameters. In those instances, we compute SNR$_{1\sigma}$ by extrapolating from subpopulations with lower SNR cutoffs. Based on our data, we find that parameter biases due to waveform errors could start becoming significant around SNR~$\geq~400$ for a typical binary from the specified population. In other words, we expect half of the detected population with SNR~$\geq~400$ to have at least one parameter that is significantly affected by waveform modeling errors.

\subsection{Estimating future accuracy requirements}
\label{subsec:future_reqs}
Waveform models with better calibration to NR, and subsequently smaller mutual mismatch across the parameter space, will be crucial to making optimal use of XG GW detections. A natural question to ask is: how much better do such waveform models need to be? One approach could be to set a benchmark for the performance of these future waveforms, say, a waveform calibration effort that results in 99\% of the expected detections to be biased by $\leq 1\sigma$ for a few chosen parameters. We can use the formalism described in Sec.~\ref{subsection:hybrid_wf} to quantify the average mismatch for which this condition will be met.

\begin{figure}[t]
\includegraphics[trim={1cm 2cm 2.7cm 3.8cm},width=0.9\linewidth]{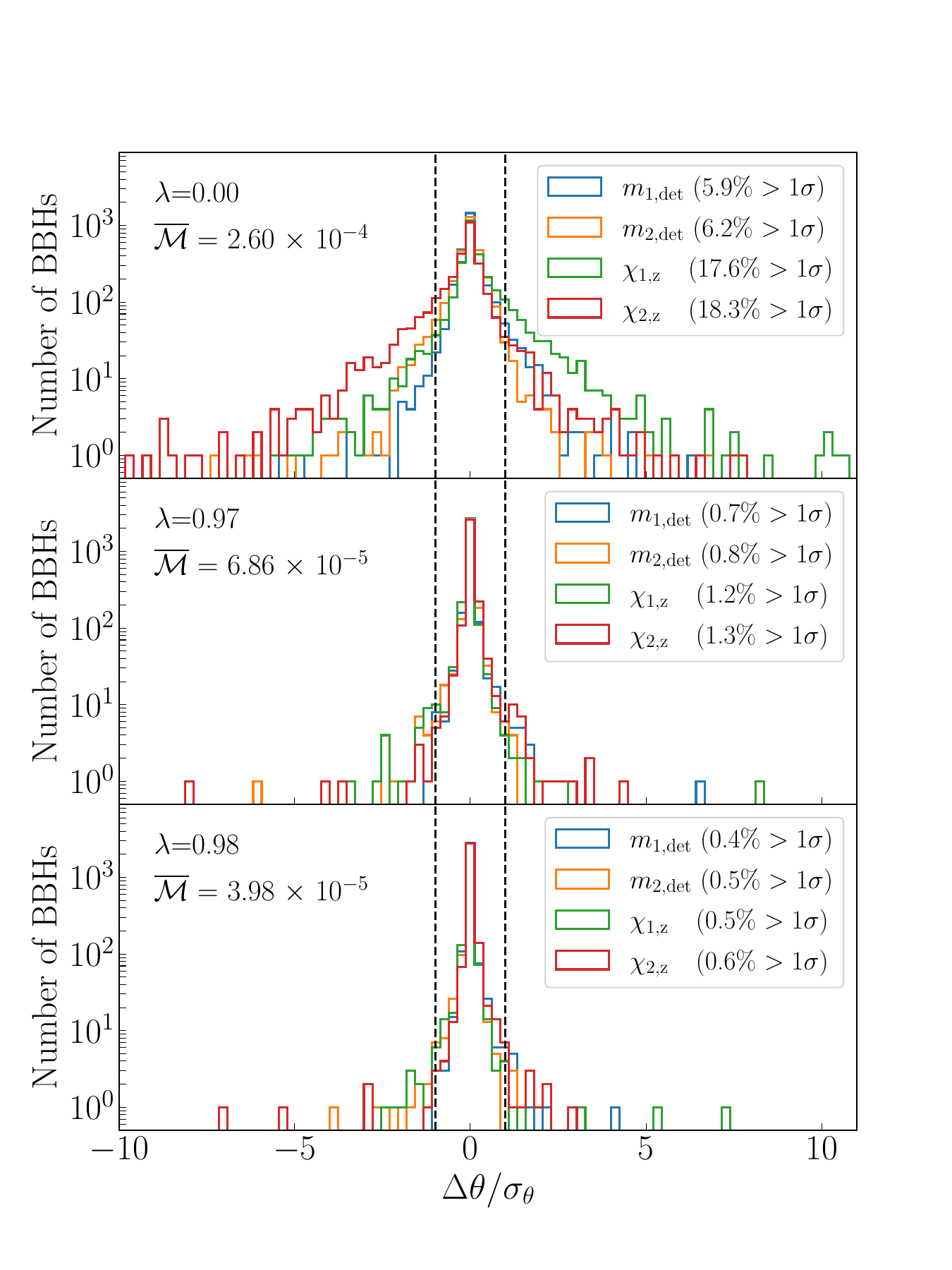}
\caption{Significance of waveform calibration biases for the SNR~$\geq~100$ subpopulation, between interpolated waveforms with varying levels of simulated accuracy to \texttt{IMRPhenomXAS}. The value of $\lambda$ used to generate the approximate waveforms and the average mismatch $\overline{\mathcal{M}}$ across the chosen subpopulation is shown in the top left corner of each panel. The dashed black lines mark the $1\sigma$ bias threshold in either direction, and the percentage of binaries with bias $> 1\sigma$ is given in the legend for each parameter. \label{fig:hybrid_bias_hists}}
\label{fig:1d_bias_hist_pop_3g}
\end{figure}

In Fig.~\ref{fig:hybrid_bias_hists}, we show the distribution of systematic biases between \texttt{IMRPhenomXAS} and various sets of interpolated waveforms with selected values of the interpolation parameter $\lambda$. The precise percentage by which each parameter is biased above the acceptable threshold depends on the specifics of the interpolated waveform, and is merely demonstrative. The general trend, however, is robust: improving the agreement between two waveforms models lowers the overall systematic bias between them. Note that biases in $D_{\rm L}$ have been omitted from the following discussion, since they are particularly sensitive to the interpolated waveforms and can be affected in unphysical ways. However, as we showed in the previous section, biases in the intrinsic parameters are typically larger than biases in $D_{\rm L}$. As such, consideration of the intrinsic parameters should be sufficient for determining the strongest constraints on the waveform accuracy threshold. See Appendix~\ref{app:interpolation} for an illustration of how the intrinsic parameters vary with $\lambda$ for a single binary.

\begin{figure*}[t!]
    \centering
        \includegraphics[width=0.96\columnwidth]{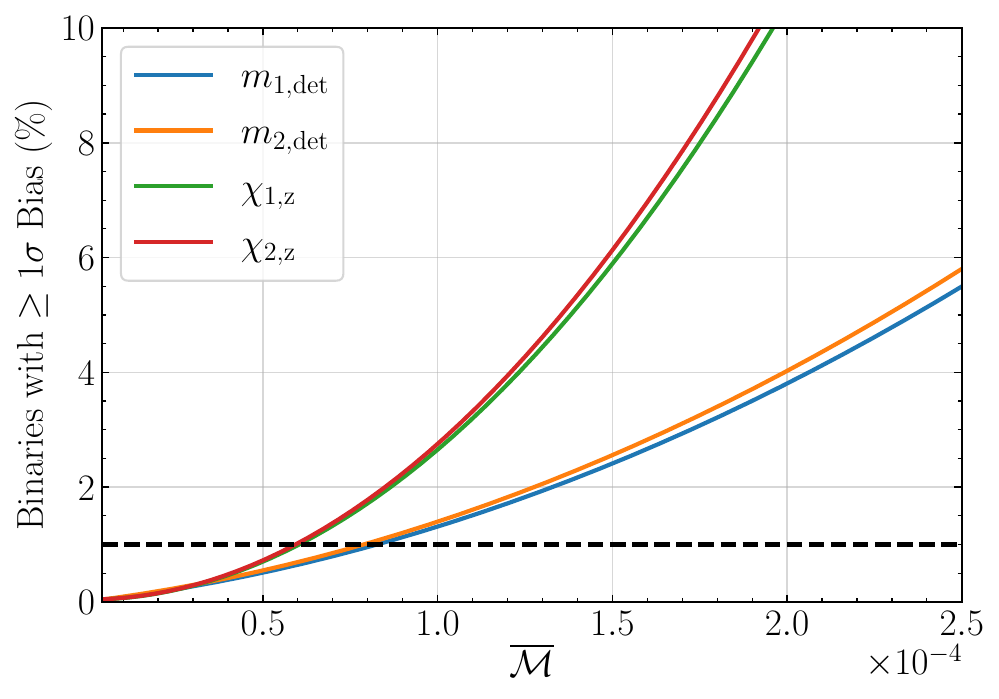}
        \includegraphics[trim={0 0.22cm 0 0},width=\columnwidth]{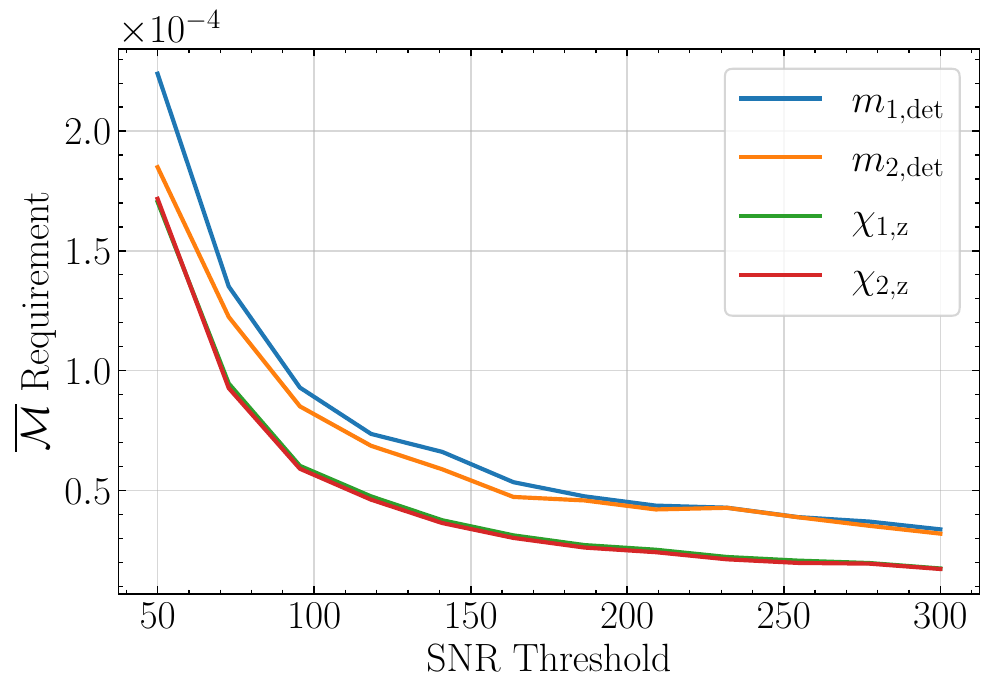}
    \caption{Mismatch requirements to have unbiased parameters, computed using the interpolated waveform formalism. Left panel: percentage of binaries that are biased by more than $1\sigma$ as a function of the average mismatch, for the SNR $\geq 100$ subpopulation. The dashed black line marks the threshold at which 1\% of events are biased by at least $1\sigma$, or equivalently, 99\% of the binaries are biased by at most $1\sigma$. Right panel: average mismatch requirement for at most 1\% of the binaries to be biased by $\geq~1\sigma$, for various SNR thresholds.}
    \label{fig:mismatch}
\end{figure*}

Performing this analysis for a finer grid of interpolated waveforms, we can develop a picture of how biased each parameter is for a range of average waveform mismatches. This result is shown in the left panel of Fig.~\ref{fig:mismatch}. As the average mismatch between waveform models decreases, the percentage of events with ${\rm SNR} \geq 100$ and bias $\geq 1\sigma$ decreases for every parameter.
In the right panel of Fig.~\ref{fig:mismatch}, we show how this mismatch requirement varies over a range of SNR cutoffs, and not just the fiducial ${\rm SNR}\geq 100$ threshold. As one considers lower SNR binaries, the average mismatch requirement for keeping 99\% of the binaries unbiased becomes less stringent across all parameters. On the other hand, the average mismatch required is much lower for subpopulations with higher SNRs. 

Based on the ${\rm SNR}\geq 100$ results, the average mismatch between the two given waveform models would need to be $\sim 6 \times 10^{-5}$ for the intrinsic parameters of 99\% of the population to remain unbiased. The mismatch requirement can be as strict as $1 \times 10^{-5}$ if we consider subpopulations with larger SNRs. Recall that the default average mismatch for the ${\rm SNR}\geq 100$ binaries was $2.6 \times 10^{-4}$. Therefore, the required mismatch is lower by nearly an order of magnitude.

\begin{figure*}[t]
\includegraphics[trim={0cm 0cm 0 0},  width=\linewidth]{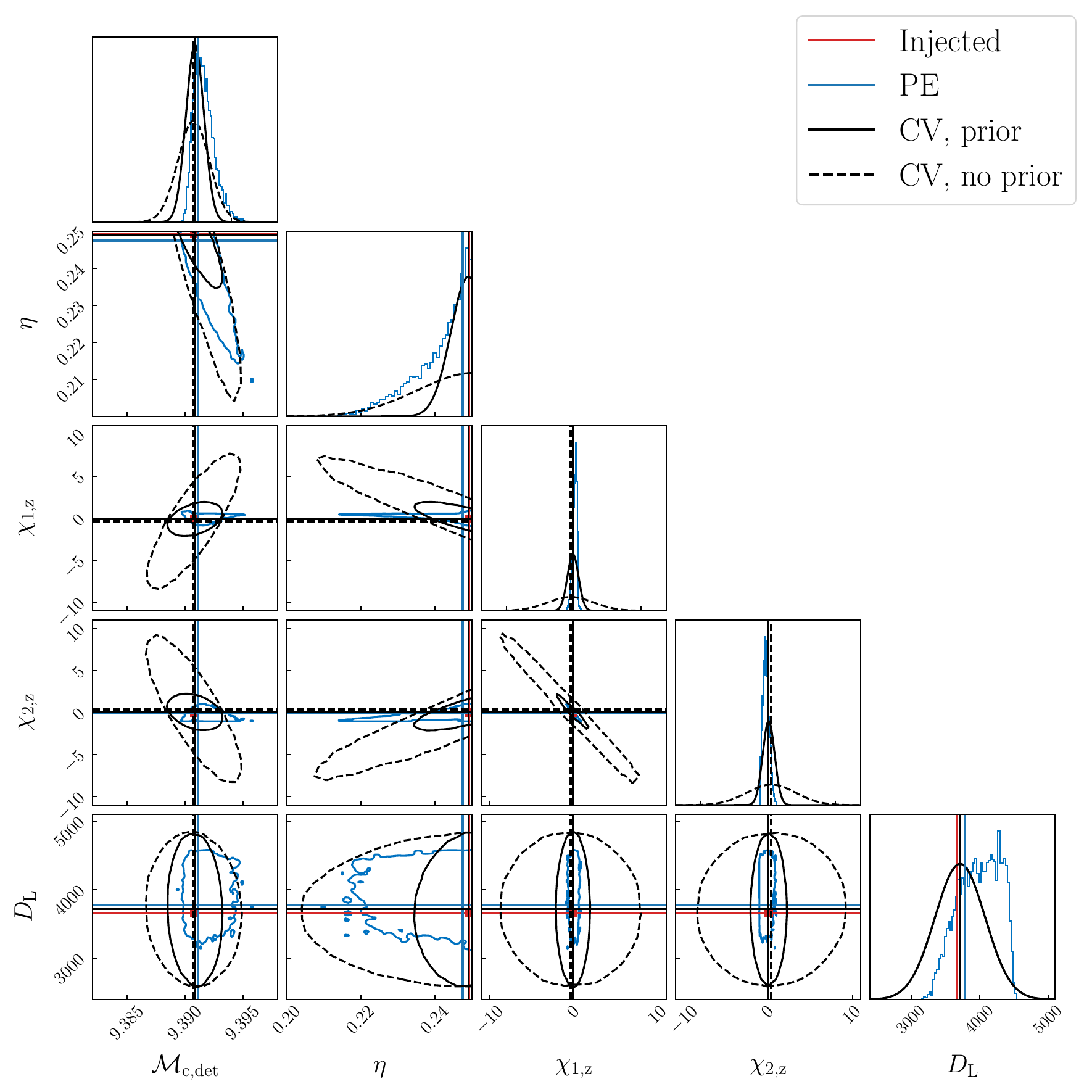}
\caption{Posterior distributions for a representative BBH in our population. The blue curves results from a full Bayesian parameter estimation run. The black curves are estimated through the Cutler-Vallisneri formalism, with (solid lines) and without (dashed lines) imposing a Gaussian priors on the spin components. The 2D contours correspond to $90\%$ confidence levels. The red lines correspond to the injected values. This particular BBH has the following true parameters: $\mathcal{M}_{\rm c, det} = 9.391\,\rm{M}_\odot$, $\eta=0.249$, $\chi_{1,\rm{z}}=-0.077$, $\chi_{2,\rm{z}}=0.047$, $D_{\rm L}=3665.092$~\rm{Mpc}, $\iota=0.631$, $\rm{RA}=0.849$, $\rm{DEC}=0.728$, $\psi=5.696$. }%
\label{fig:typical_pe}
\end{figure*}

\begin{figure*}[t]
\includegraphics[trim={0cm 0cm 0 0},  width=\linewidth]{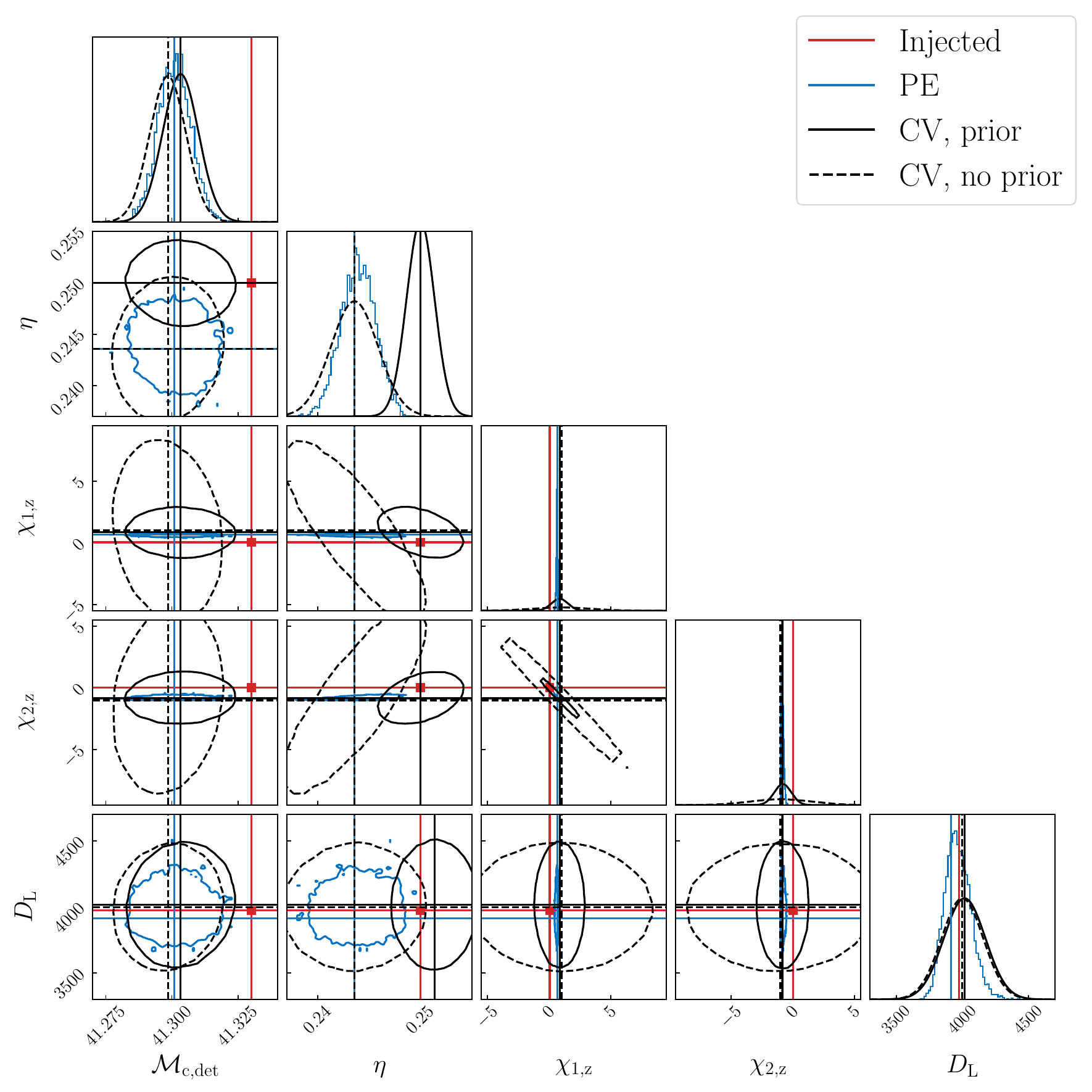}
\caption{Same as Fig.~\ref{fig:typical_pe}, but now for a pessimistic case where the intrinsic parameters recovered from parameter estimation are strongly biased. The BBH has the following true parameters: $\mathcal{M}_{\rm c, det} = 41.330\,\rm{M}_\odot$, $\eta=0.250$, $\chi_{1,\rm{z}}=0.050$, $\chi_{2,\rm{z}}=0.024$, $D_{\rm L}=3972.875$~\rm{Mpc}, $\iota=0.733$, $\rm{RA}=2.680$, $\rm{DEC}=-0.516$, $\psi=0.868$. %
} %
\label{fig:pessimistic_pe}
\end{figure*}

This statement is sensitive to several assumptions, including the choice of population, the waveform models in question, and even the parameters for which the bias should be minimized. Also, we stress that this requirement applies to the {\em average} mismatch across our population, and not to individual binaries. Many low-SNR events in specific regions of the parameter space may require less demanding waveform accuracy, while louder binaries may require even better waveform calibration. However, the approach adopted here provides a reasonable and conservative order-of-magnitude estimate for the mismatch requirement across an expected population. 

\section{Comparison with full parameter estimation}
\label{sec:fullPE}

We now validate our Fisher results and Cutler-Vallisneri estimates by comparing them with full Bayesian parameter estimation runs using \texttt{BILBY}~\cite{Ashton:2018jfp,Krishna:2023bug}. To mimic the biases from imperfect waveform modeling, we inject a GW signal with the \texttt{IMRPhenomXAS} model and recover it with the \texttt{IMRPhenomD} model.

In Fig.~\ref{fig:typical_pe} we show a representative case belonging to the subpopulation of binaries with SNR $\geq 100$.
This particular BBH has a network SNR of $\sim140$. We compare the parameter estimation results against Gaussian posteriors centered at the biased value $\theta_{i, \rm biased}=\theta_i+\Delta\theta_i$, where $\theta_i$ are the injected parameters, $\Delta\theta_i$ is the Cutler-Vallisneri bias, and the covariance matrix is the inverse of the Fisher matrix. Solid (dashed) black lines refer to the case in which we do (or do not) impose Gaussian priors on the spin components following the simple recipe of Eq.~\eqref{eq:prior}.

We find that the biased estimates from Cutler-Vallisneri, $\theta_{i, \rm biased}$, for all parameters relevant to this study are close to
the best-estimate parameters recovered with \texttt{BILBY}, whether we do or do not impose priors. The Fisher errors $\sigma_{i, \rm stat}$, instead, tend to be larger than the corresponding intervals in the $1$D marginalized posteriors. For $\mathcal{M}_{\rm c, det}$, $\eta$, and $D_{\rm L}$ (or equivalently $m_{1,\rm det}$, $m_{2,\rm det}$, and $z$) the discrepancy between Fisher errors and $1\sigma$ intervals in the marginalized posteriors is within a factor of $\lesssim 2$.
If we do not impose a prior on the spins, the Fisher errors on $\chi_{1,\rm z}$ and $\chi_{2,\rm z}$ (dashed black contours) are larger by more than an order of magnitude than the corresponding confidence intervals in the $1$D marginalized posteriors from \texttt{BILBY}. When we impose a Gaussian prior (solid black contours), the disagreement decreases to a factor of $\lesssim 3$ for both spin components. However, due to parameter degeneracies, imposing a prior on the spins also affects $\mathcal{M}_{\rm c, det}$ and $\eta$. For $\mathcal{M}_{\rm c, det}$, the Fisher error goes from being $\sim 50\%$ larger than the $1\sigma$ interval in the marginalized posteriors to being $\sim 15\%$ smaller. For $\eta$, the Fisher error goes from $\sim 2$ times larger to $\sim 40\%$ smaller.

The 
discrepancy in the errors %
might lead us to underestimate the statistical significance of the biases with the Cutler-Vallisneri formalism. To illustrate how this may happen, in Fig.~\ref{fig:pessimistic_pe} we compare Fisher estimates and parameter estimation results for the ``pessimistic'' case of a BBH for which the recovered parameters are significantly biased.
The corresponding signal has a network SNR of $\sim 350$. For this BBH, the Cutler-Vallisneri biases and the Fisher matrix estimates of the errors on $\mathcal{M}_{\rm c, det}$, $\eta$, and $D_{\rm L}$ are in reasonable agreement with the best-estimate parameters from full parameter estimation without imposing any Gaussian prior. The Fisher errors on these parameters agree within $\lesssim 50\%$ with the $1\sigma$ intervals in the $1$D marginalized posteriors. However, the Fisher matrix estimates of the errors on $\chi_{1,\rm{z}}$ and $\chi_{2,\rm{z}}$ are larger than the parameter estimation results by a factor $\sim 40$ without any prior, while the Cutler-Vallisneri biases would far exceed the physical range, thus we force them to take the boundary values $\theta_{\chi_{i,\rm{z}}, \rm biased} \in [-1, 1]$. %
This is a clear example of why the spin priors are necessary. After imposing the prior, the Cutler-Vallisneri biases agree with the best estimate from parameter estimation within $\sim 30\%$ for both spin components, and the $1\sigma$ errors are now larger by a factor of $\sim 10$ instead of $\sim 40$. Still, the spin components recovered with full parameter estimation are both biased at $\sim10\sigma$ with respect to the injected values, while they are barely biased (at $\sim1\sigma$) when we use the Cutler-Vallisneri formalism. Moreover, while the $1\sigma$ Fisher errors on the marginalized distributions for $\mathcal{M}_{\rm c, det}$, $\eta$ and $D_{\rm L}$ are only slightly affected by the imposition of a spin prior, the Cutler-Vallisneri bias on $\eta$ changes significantly, now pushing this parameter to hit the physical boundary of $0.25$.

In general, we find that the combination of Fisher and Cutler-Vallisneri formalism might perform poorly for bounded parameters when the estimated uncertainty exceeds the prior bound, as in Fig.~\ref{fig:pessimistic_pe}.
A more systematic comparison of the Cutler-Vallisneri formalism with full parameter estimation in the whole parameter space is clearly necessary, but it is beyond the scope of this exploratory study.

\section{Discussion}
\label{sec:discussion}

Our estimates suggest that, in order for 99\% of the intrinsic parameters of an astrophysical BBH population with SNR~$\geq~100$ to remain unbiased by waveform errors, an order of magnitude improvement in the average mismatch is required. Our estimate is meant to be a lower bound, especially when considered alongside other work where the authors study individual binaries with SNRs of the order $\mathcal{O}(1000)$~\citep{Purrer:2019jcp,Hu:2022rjq}. We stress that the mismatch and waveform calibration requirements will ultimately depend on the intended application. Over a population of binaries, many systems will have insignificant biases in intrinsic parameters due to either lower SNRs or the existing calibration accuracy of specific waveform models. For the loudest binaries observed in XG detectors, the biases will be extremely significant, which may be particularly relevant for applications such as dark siren cosmology or tests of general relativity.

One relevant extension of this work would be to investigate how waveform systematics over a population could impact the inference of specific hyperparameters of a given population model. The accuracy requirements for ensuring unbiased population hyperparameters may differ from the results presented here in nontrivial ways.

Additionally, the waveform models used in this study both belong to the \texttt{IMRPhenom} family, and represent only a subset of the models that may be used in a parameter estimation campaign. Models with higher harmonics, spin precession, etc. could produce different mutual biases, and would likely require higher accuracy thresholds to ensure consistent parameter inference. The interpolated waveform formalism presented in this work (or generalizations thereof) could be especially useful in guiding the calibration efforts for these models. By varying waveform accuracy in specific parts of the parameter space, one could deduce where to iteratively add BBH simulations to most efficiently improve waveform accuracy with additional simulations. These applications are left for future studies.

\acknowledgments

The authors are thankful to A. Dhani, S. V{\"o}lkel, and A. Buonanno for pointing out the importance of waveform alignment, as well as for generously helping us in implementing the alignment procedure in our code. We also thank Ssohrab Borhanian, Ken Ng, Andrea Antonelli, Carl-Johan Haster, Nico Yunes, and Bore Gao for helpful discussions.
V.K., R.C., L.R. and E.B. are supported by NSF Grants No. AST-2006538, No. PHY-2207502, No. PHY-090003, and No. PHY-20043, by NASA Grants No. 20-LPS20-0011 and No. 21-ATP21-0010, by the John Templeton Foundation Grant 62840, by the Simons Foundation, and by the Italian Ministry of Foreign Affairs and International Cooperation Grant No.~PGR01167.
This work was carried out at the Advanced Research Computing at Hopkins (ARCH) core facility (\url{rockfish.jhu.edu}), which is supported by the NSF Grant No.~OAC-1920103.

\appendix

\section{Waveform biases in a detector network}
\label{app:network_cv_bias}

In this appendix we generalize the calculation of systematic parameter biases due to waveform errors from Ref.~\cite{Cutler:2007mi} to a network of detectors.
Let us assume that a GW data stream can be modeled as a combination of some true gravitational waveform and noise, such that 
\begin{equation}
    d(t) = h_{\rm TR}(t; \vec{\theta}_{\rm tr}\,) + n(t),
    \label{eq:data_stream}
\end{equation}
where the signal is a function of time $t$, and $\vec{\theta}_{\rm tr}$ represents the set of true parameter values that we would like to infer. In practice, this is done using an approximate waveform model $h_{\rm AP}$, such that
\begin{equation}
    h_{\rm TR}(t; \vec{\theta}_{\rm tr}\,) \neq h_{\rm AP}(t; \vec{\theta}_{\rm tr}\,).
\end{equation}
Parameter estimation involves minimizing the following log-likelihood with respect to $\vec{\theta}$~\citep{LIGOScientific:2019hgc}:
\begin{equation}
    \log p(d|\vec{\theta}) = -\frac{1}{2} \left( d(t) - h_{\rm AP}(t;\vec{\theta}\, ) \mid d(t) - h_{\rm AP}(t;\vec{\theta}\, )\right).
\end{equation}
In the presence of multiple detectors, the various antenna patterns and sky locations will give rise to detector-dependent waveforms, such that parameter estimation requires minimizing the following log-likelihood instead:
\begin{multline}
    \log p(d|\vec{\theta}) = \\
    \Sigma_{\rm D=1}^{N_{\rm D}} -\frac{1}{2} \left( d_{\rm D}(t) - h_{\rm AP, D}(t;\vec{\theta}\, ) \mid d_{\rm D}(t) - h_{\rm AP, D}(t;\vec{\theta}\,)\right).
\end{multline}
The best-fit parameters $\vec{\theta}_{\rm bf}$ will thus satisfy
\begin{equation}
    \Sigma_{\rm D=1}^{N_{\rm D}} \left( \partial_i h_{\rm AP, D}(\vec{\theta}_{\rm bf}) \mid  d_{\rm D}(t) - h_{\rm AP, D}(\vec{\theta}_{\rm bf})\right) = 0,
    \label{eq:bf_likelihood_condition}
\end{equation}
where here and below we omit the time dependence for brevity, and the index $i$ represents the components of $\vec{\theta}_{\rm bf}$.
Using Eq.~\eqref{eq:data_stream}, we can write 
\begin{multline}
    d_{\rm D}(t) - h_{\rm AP, D}(\vec{\theta}_{\rm bf}\,) = n_{\rm D}(t) + h_{\rm TR, D}( \vec{\theta}_{\rm tr}) \\
    - h_{\rm AP, D}(\vec{\theta}_{\rm tr})
    + h_{\rm AP, D}(\vec{\theta}_{\rm tr}) - h_{\rm AP, D}(\vec{\theta}_{\rm bf}),
    \label{eq:log_like_expansion}
\end{multline}
where we have added and subtracted a term $h_{\rm AP, D}(\vec{\theta}_{\rm tr})$ for convenience. We can think of $\vec{\theta}_{\rm bf}$ as a perturbation around the true parameters, such that
\begin{equation}
    \vec{\theta}_{\rm bf} = \vec{\theta}_{\rm tr} + \Delta \vec{\theta}.
\end{equation}
If $\Delta \vec{\theta}$ is sufficiently small, the linear signal approximation gives us
\begin{equation}
    h_{\rm AP, D}(\vec{\theta}_{\rm tr}) - h_{\rm AP, D}(\vec{\theta}_{\rm bf}) \approx \partial_i h_{\rm AP, D}(\vec{\theta}_{\rm bf}) \Delta \theta^i.
\end{equation}
Therefore, we can rewrite Eq.~\eqref{eq:log_like_expansion} as
\begin{multline}
    d_{\rm D}(t) - h_{\rm AP, D}(\vec{\theta}_{\rm bf}\,) \approx \\
    n_{\rm D}(t) + \delta h_{\rm D}(\vec{\theta}_{\rm tr}) - \partial_i h_{\rm AP, D}(\vec{\theta}_{\rm bf}) \Delta \theta^i,
\end{multline}
where $\delta h_{\rm D}(\vec{\theta}_{\rm tr}) \equiv h_{\rm TR, D}(\vec{\theta}_{\rm tr}) - h_{\rm AP, D}( \vec{\theta}_{\rm tr})$ is the waveform residual.

Substituting this expansion back into Eq.~\eqref{eq:bf_likelihood_condition} and redefining some indices, we obtain
\begin{dmath}
    \Sigma_{\rm D=1}^{N_{\rm D}} \left( 
    \partial_i h_{\rm AP, D}(\vec{\theta}_{\rm bf}) \mid  n_{\rm D}(t) + 
    \delta h_{\rm D}(\vec{\theta}_{\rm tr}) - \partial_j h_{\rm AP, D}(\vec{\theta}_{\rm bf}) \Delta \theta^j 
    \right) = 0.
\end{dmath}
Expanding the inner product terms, we get
\begin{equation}
\begin{split}
    \Sigma_{\rm D=1}^{N_{\rm D}} &\left( \partial_i h_{\rm AP, D}(\vec{\theta}_{\rm bf}) \mid  n_{\rm D}(t) \right) \\ 
    + \Sigma_{\rm D=1}^{N_{\rm D}} &\left( \partial_i h_{\rm AP, D}(\vec{\theta}_{\rm bf}) \mid  \delta h_{\rm D}(\vec{\theta}_{\rm tr}) \right) \\
    &= \Sigma_{\rm D=1}^{N_{\rm D}} \left( \partial_i h_{\rm AP, D}(\vec{\theta}_{\rm bf}) \mid \partial_j h_{\rm AP, D}(\vec{\theta}_{\rm bf}) \right)\Delta \theta^j
\end{split}
\end{equation}
The inner product on the right-hand side can be recognized as the network Fisher information matrix of $h_{\rm AP, D}$, such that
\begin{equation}
\begin{split}
    \Sigma_{\rm D=1}^{N_{\rm D}} &\left( \partial_j h_{\rm AP, D}(\vec{\theta}_{\rm bf}) \mid  n_{\rm D}(t) \right) \\ 
    + \Sigma_{\rm D=1}^{N_{\rm D}} &\left( \partial_j h_{\rm AP, D}(\vec{\theta}_{\rm bf}) \mid  \delta h_{\rm D}(\vec{\theta}_{\rm tr}) \right) \\
    &= \Gamma_{ij, \rm Net} \Delta \theta^i,
\end{split}
\end{equation}
where we have used the symmetry of the Fisher matrix and switched the $i, j$ indices for convenience.
The overall bias on parameter $\theta^i$ can now be expressed as
\begin{dmath}
    \Delta \theta^i = (\Gamma^{-1}_{\rm Net})^{ij}(\vec{\theta}_{\rm bf}) \left[ \Sigma_{\rm D=1}^{N_{\rm D}} \left( \partial_j h_{\rm AP, D}(\vec{\theta}_{\rm bf}) {\mid}  n_{\rm D}(t) \right) \\
    + \Sigma_{\rm D=1}^{N_{\rm D}} \left( \partial_j h_{\rm AP, D}(\vec{\theta}_{\rm bf}) {\mid} \delta h_{\rm D}(\vec{\theta}_{\rm tr}) \right) \right].
\end{dmath}

Each of the two terms on the right-hand side has a clear interpretation: the first term represents the error contribution from noise, while the second is the systematic bias due to waveform residuals. Isolating the systematic waveform error term, we can write
\begin{dmath}
    \Delta \theta^i_{\rm sys} \equiv (\Gamma^{-1}_{\rm Net})^{ij}(\vec{\theta}_{\rm bf}) \Sigma_{\rm D=1}^{N_{\rm D}} \left( \partial_j h_{\rm AP, D}(\vec{\theta}_{\rm bf}) {\mid} \delta h_{\rm D}(\vec{\theta}_{\rm tr}) \right).
\end{dmath}
Since in practice we do not have access to $\vec{\theta}_{\rm tr}$, we can instead evaluate the waveform residual at $\vec{\theta}_{\rm bf}$ to a good approximation. This yields the final result:
\begin{dmath}
    \Delta \theta^i_{\rm sys} = (\Gamma^{-1}_{\rm Net})^{ij}(\vec{\theta}_{\rm bf}) \Sigma_{\rm D=1}^{N_{\rm D}} \left( \partial_j h_{\rm AP, D}(\vec{\theta}_{\rm bf}) {\mid} h_{\rm TR, D}(\vec{\theta}_{\rm bf}) - h_{\rm AP, D}(\vec{\theta}_{\rm bf}) \right).
\end{dmath}

\section{Impact of waveform alignment on luminosity distance biases}
\label{app:alignment}

Systematic errors due to different waveforms are of the form
\begin{equation}
    \Delta \theta^i =  (\Gamma_{\rm AP, Net}^{-1})^{ij} \, (\partial_j (h_{\rm AP})| h_{\rm TR} - h_{\rm AP}).
\end{equation}
In the frequency domain, the overall amplitudes of $h_{\rm TR}$ and $h_{\rm AP}$ are independent of systematic shifts in phase and time. However, differences in the coalescence time and phase affect the relative phases of the two waveforms. Since the Cutler-Vallisneri bias formula depends on the difference $(h_{\rm TR}-h_{\rm AP})$, the biases are also affected by the relative phase between the two waveform models.

\begin{figure}[t]
\includegraphics[trim={0cm 0cm 0cm 0}, 
width=\linewidth]{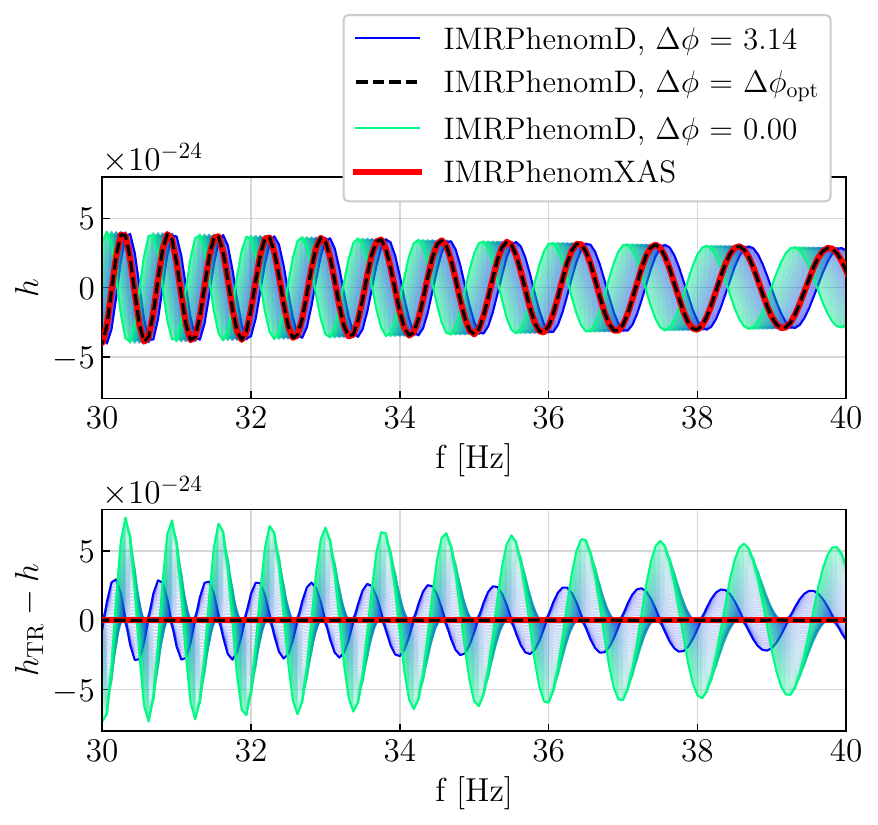}

\caption{Top: GW strain amplitude for a chosen binary when using various waveform models. The reference waveform generated using \texttt{IMRPhenomXAS} is shown in red. We also show different versions of an \texttt{IMRPhenomD} waveform with the same injected parameters and an additional phase shift in the range [$0$,$\pi$] (curves ranging from green to blue). The waveform which minimizes the mismatch with respect to the reference waveform is plotted as a dashed black curve.
Bottom: the same set of waveforms, with the y-axis now showing the difference between the given waveform and the \texttt{IMRPhenomXAS} reference waveform.}
\label{fig:dl_vs_phic}
\end{figure}

In Fig.~\ref{fig:dl_vs_phic} we illustrate the effect of changing one of these parameters, $\phi_{\rm c}$. As shown in the bottom panel, the difference between the two waveforms can depend strongly on their relative phase difference. Biases in luminosity distance have the primary effect of changing the relative amplitude of the template waveform, which can cancel out this effect. Therefore, biases in luminosity distance from the Cutler-Vallisneri formalism can be unphysically large if the waveforms are not aligned first.

As such, the alignment procedure alluded to in Sec.~\ref{sec:populations} has a dramatic impact on the bias in $D_{\rm L}$.

\section{Choice of reference frequency}
\label{app:fref}

\begin{figure}[t]
\includegraphics[trim={0cm 0.5cm 0cm 0}, 
width=\linewidth]{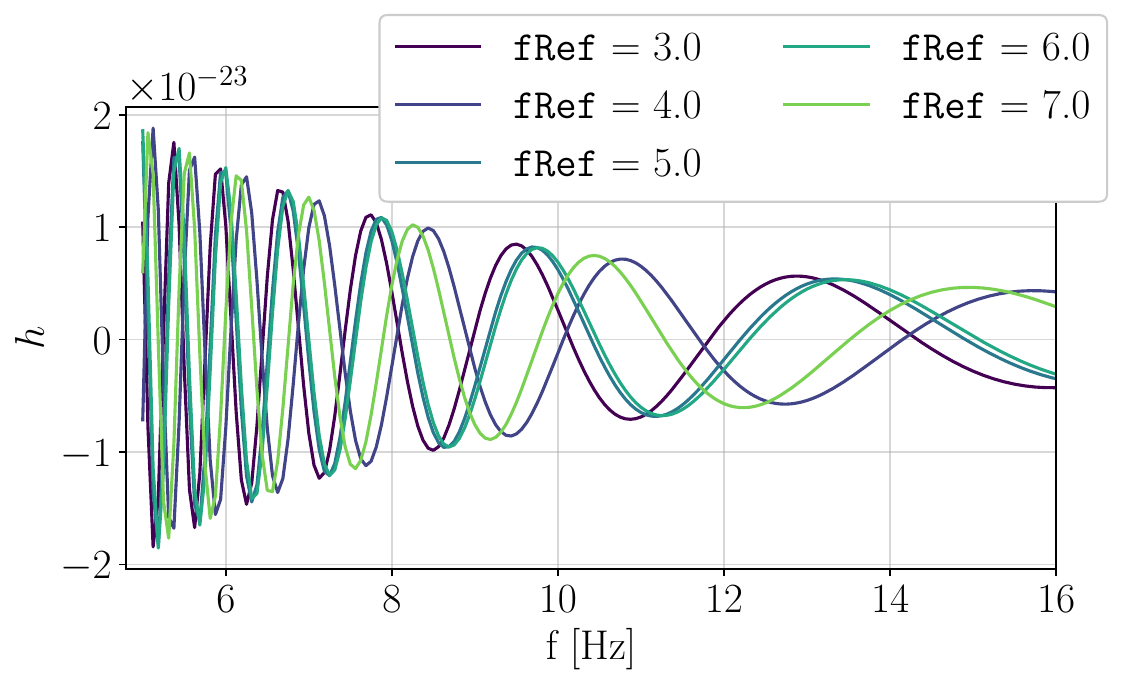}
\caption{Detector response of a given BBH from our population, for various choices of \texttt{fRef}. The detector is set to the 40 km CE in Idaho, and the BBH has the following parameters: $\mathcal{M}_{\rm c, det}=121.181~\rm{M}_\odot$, $\eta=0.249$,  $\chi_{1, \rm{z}}=0.026$, $\chi_{2, \rm{z}}=0.036$, $D_{\rm L}= 16215.376$~Mpc, $\iota=2.709$, ${\rm RA}=1.511$, ${\rm DEC}=0.188$, $\psi=2.049$.}
\label{fig:fRef_dep_wf}
\end{figure}

The waveforms for this study are generated using the \texttt{LALSuite} library~\citep{lalsuite, swiglal}, which requires a parameter ``\texttt{fRef}'' to set the reference frequency at which the phase and orientation of a binary are defined. For this study we set \texttt{fRef} to 5 Hz, which is the minimum frequency chosen for the detector network. In Fig.~\ref{fig:fRef_dep_wf} we show the effect of varying \texttt{fRef} while keeping all other parameters constant, for a specific binary. 

Evidently, the choice of \texttt{fRef} can affect the behavior of a frequency domain waveform in much the same way as an additional phase factor, and in this sense it has a similar effect as modifying the coalescence phase parameter $\phi_{\rm c}$.

\begin{figure}[h]
\includegraphics[trim={1.6cm 0.5cm 0.3cm 0}, 
width=0.9\linewidth]{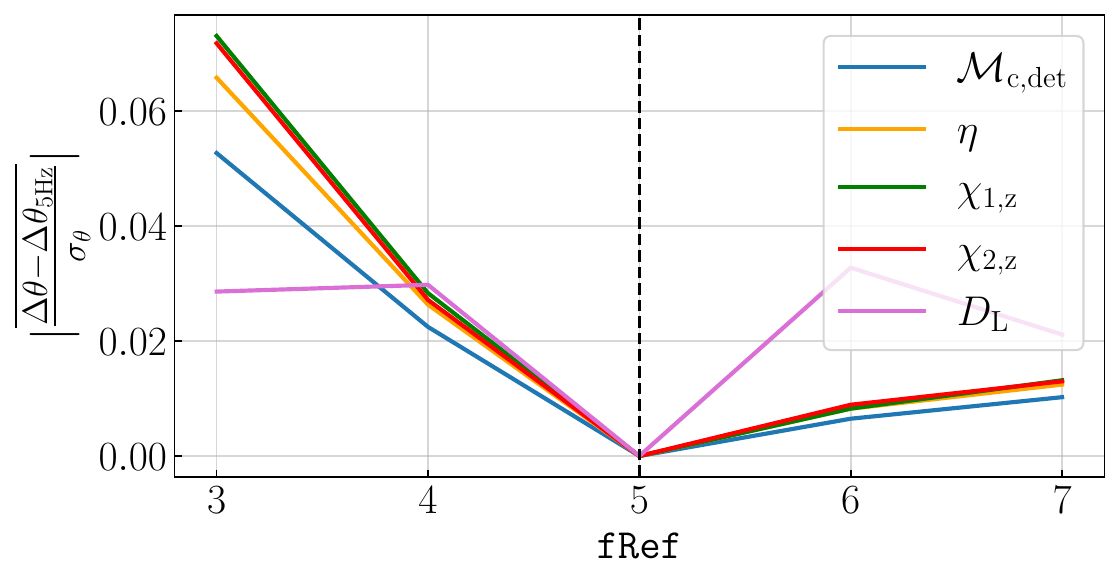}
\caption{Average deviations in parameter biases over a subset of 100 binaries with ${\rm SNR} \geq 100$ from our population, for various choices of \texttt{fRef}. The y-axis shows the deviation of the Cutler-Vallisneri biases with respect to their respective values at \texttt{fRef}=5~Hz, normalized to the statistical errors for each binary. The dashed black line marks the location of \texttt{fRef}=5~Hz.}
\label{fig:fRef_dep}
\end{figure}

Although the waveform alignment process alluded to in Sec.~\ref{sec:populations} should generally account for variations in the waveform phase, there may still be some minor effects on the Cutler-Vallisneri bias as a result of different choices of \texttt{fRef}. The tests shown in Fig.~\ref{fig:fRef_dep} demonstrate that the choice of \texttt{fRef} has a minor impact on the parameter biases. Using the biases at \texttt{fRef}=5~Hz as a reference, changing the reference frequency only changes the normalized biases by a few parts in a hundred.

\section{Biases using interpolated waveforms}
\label{app:interpolation}

\begin{figure}[h]
\includegraphics[trim={1.0cm 0.cm 1.0cm 0}, 
width=0.9\linewidth]{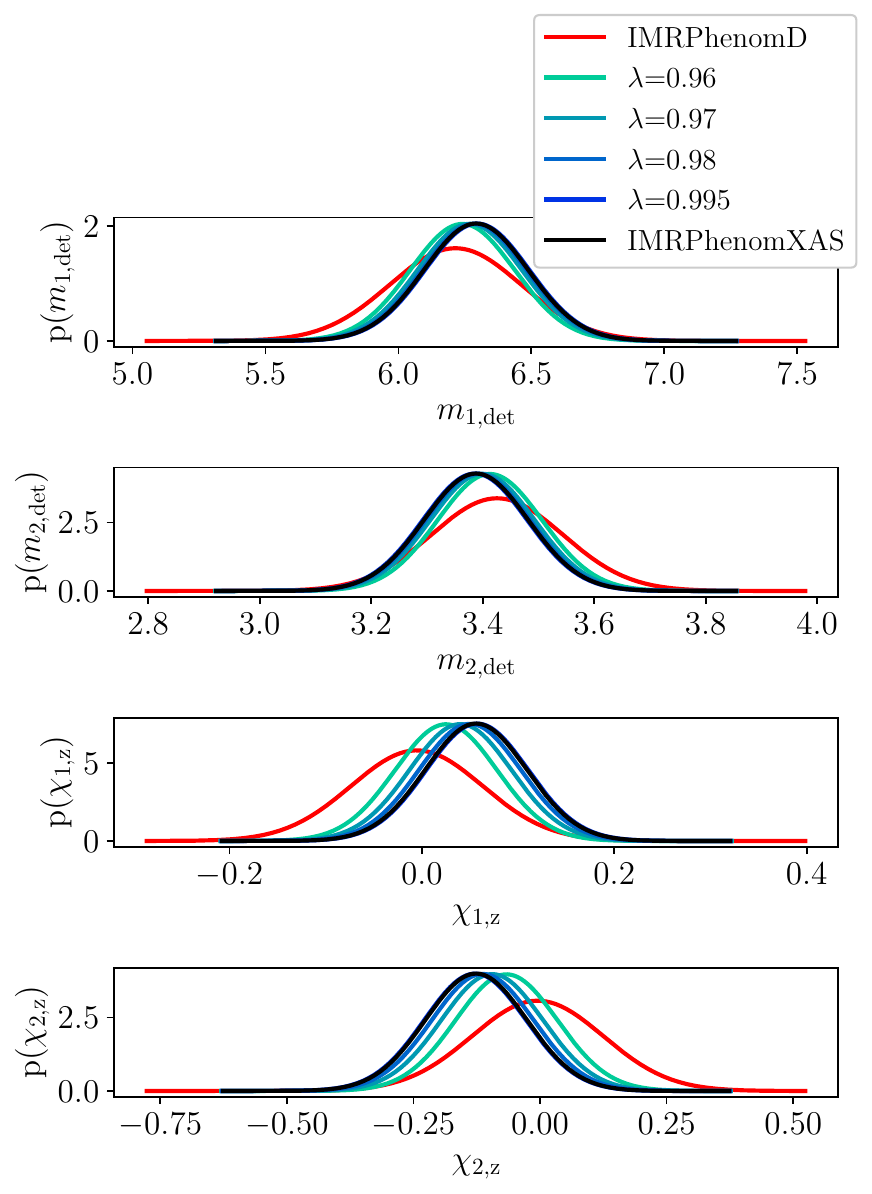}
\caption{Behavior of statistical and systematic errors for a single BBH event, for various interpolated waveforms defined by the parameter $\lambda$. The chosen BBH has the following parameters: $\mathcal{M}_{\rm c, det}=3.982~\rm{M}_\odot$, $\eta=0.227$,  $\chi_{1, \rm{z}}=0.056$, $\chi_{2, \rm{z}}=-0.126$, $D_{\rm L}= 131.168$~Mpc, $\iota=2.162$, ${\rm RA}=5.575$, ${\rm DEC}=-0.654$, $\psi=2.663$.}
\label{fig:hybr_test_binary}
\end{figure}

Using the interpolation scheme described in Sec.~\ref{subsection:hybrid_wf}, we synthetically produce waveforms that vary monotonically in mismatch between \texttt{IMRPhenomD} and \texttt{IMRPhenomXAS}. In Fig.~\ref{fig:hybr_test_binary} we illustrate the modulation of the statistical and systematic uncertainties from the Fisher and Cutler-Vallisneri formalism, by focusing on a single test binary. As expected, the magnitude of the biases as well as the statistical uncertainties of the intrinsic parameters can be seen changing from the value in \texttt{IMRPhenomD} ($\lambda=0$) to the value in \texttt{IMRPhenomXAS} ($\lambda=1$). This example shows that the interpolated waveform formalism can be trusted to produce qualitatively sensible results for the intrinsic parameters considered in Sec.~\ref{subsec:future_reqs}.

\bibliography{refs_prd}

\end{document}